\documentclass[preprint,review,11pt,authoryear]{elsarticle}

\usepackage{amsmath,amssymb}
 \usepackage{amsthm}
 \usepackage{bbm}
 \usepackage{graphicx}
\usepackage{color}
\definecolor{cornell-red}{RGB}{179,27,27}
\usepackage{subcaption}
\usepackage{mathtools}
\usepackage{dsfont}
\usepackage{color}
\usepackage{natbib}
    \usepackage{multirow}
\usepackage{pgfplots}
\usepackage{caption,subcaption}
\usepackage[left=1in,right=1in,top=1in,bottom=1in]{geometry}
\usepackage{indentfirst}

\usepackage{xcolor}
\usepackage[colorlinks = true,linkcolor = blue,urlcolor  = blue,citecolor = blue,anchorcolor = blue]{hyperref}

\usepackage{booktabs}

\usepackage{natbib}

\graphicspath{ {Figures/} {Figures New/} }

\theoremstyle{definition}

\theoremstyle{remark}
\newtheorem{rem}{\textbf{Remark}}

\newcommand{\calX}{\mathcal{X}}

\usepackage{titlesec}

    \titlespacing{\section}{0pt}{0ex}{0ex}
    \titlespacing{\subsection}{0pt}{0ex}{0ex}
    \titlespacing{\subsubsection}{0pt}{0ex}{0ex}
    
    \journal{arXiv}
    \date{ }
\begin{document}

\begin{frontmatter}

\title{Equity in Stochastic Healthcare Facility Location}

\author[mymainaddress1]{Karmel S. Shehadeh\corref{cor1}}
\cortext[cor1]{Corresponding author. }
 \ead{kas720@lehigh.edu; kshehadeh@lehigh.edu}
\author[mymainaddress1]{Lawrence V. Snyder}

 \address[mymainaddress1]{Department of Industrial and Systems Engineering, Lehigh University, Bethlehem, PA,  USA}

\begin{abstract}

\noindent  We consider issues of equity in stochastic facility location models for healthcare applications. We explore how uncertainty exacerbates inequity and examine several equity measures that can be used for stochastic healthcare location modeling. We analyze the limited literature on this subject and highlight areas of opportunity for developing tractable, reliable, and data-driven approaches that might be applicable within and outside healthcare operations. Our primary focus is on exploring various ways to model uncertainty, equity, and facility location, including modeling aspects (e.g., tractability and accuracy) and outcomes (e.g., equity/fairness/access performance metrics vs. traditional metrics like cost and service levels).


\begin{keyword} 
Equity, Healthcare, Facility Location, Uncertainty, Inequity-averse Optimization, Stochastic Optimization
\end{keyword}

\end{abstract}
\end{frontmatter}

\section{Introduction}\label{sec1}

\noindent \textit{Equity} concerns naturally arise in many real-life applications (e.g., healthcare scheduling, facility location, disaster response operations, air traffic control, etc.), and it is crucial to address these concerns for the proposed solutions to be applicable, equitable, and acceptable. However, accounting for equity is a complex task primarily because there is no unique notion of equity that is generally accepted; the definition of equity often depends on the application. Moreover, uncertainty, an intrinsic property of real-life applications, interacts with equity in complicated and poorly understood ways. In particular, uncertainty increases the complexity of quantifying the loss in efficiency (e.g., minimizing the operational cost of vaccination clinics) incurred in the pursuit of equity (e.g., equitably allocating mobile vaccination clinics across geographical areas). In addition, deterministic models of most real-world problems are often challenging. Thus, incorporating both uncertainty and equity metrics or constraints  may increase the complexity of these problems.

In this chapter, we consider issues of equity in stochastic facility location models for healthcare applications. The main motivation behind the attempt to establish equity in healthcare in general---and the need for inequity-averse models for healthcare facility (HCF) location in particular---is, of course, an ethical one: \textit{humans have equal rights, and therefore, nobody should be discriminated against by inequitable access to healthcare services or distribution of healthcare services}.  Another more pragmatic motivation for striving for equitable HCF location--allocation solutions is \textit{the need to avoid adverse health outcomes of vulnerable populations who often do not have proper or equal access to HCF} \citep{gutjahr2018equity}. Finally, extending classical models with an emphasis on equity and equity--uncertainty interaction is practically relevant and technically interesting.

We first analyze different aspects and measures of equity in the literature. Then, we analyze recent static and mobile HCF location models to explore various ways to model uncertainty, equity, and facility location, including modeling aspects (e.g., tractability and accuracy) and outcomes (e.g., equity-,  fairness-, or access-based performance metrics vs.\ traditional metrics like cost and service levels).  Our goal is not to provide a comprehensive survey; instead, our goal is to highlight the issue of equity and access and the need for data-driven and tractable models to address emerging stochastic HCF problems. 

The remainder of this chapter is organized as follows: In Section~\ref{Sec2:Equity}, we analyze existing definitions and metrics of equity as well as common methods to model these metrics.  In Section~\ref{Sec3:EquityVsUncer}, we briefly discuss the challenges of incorporating uncertainty and equity and demonstrate through a simple example that uncertainty and equity interact in ways that should not be ignored. In Section~\ref{sec:EquityInHCF}, we provide a high-level analysis of recent stochastic HCF location literature with a particular focus on studies that proposed and analyzed inequity-averse approaches. Maybe not surprisingly, and sadly, this analysis reveals that there is limited literature considering equity and equity--uncertainty interaction.  Nevertheless, there are many opportunities to use the powerful tools of operations research to address equity concerns in emerging HCF location problems and derive inequity-averse stochastic HCF location approaches. In Section~\ref{Sec:Future}, we present some future research opportunities and open questions.


\section{What is Equity, Anyway?}\label{Sec2:Equity}

\noindent To account for \textit{equity} (often synonymously called \textit{fairness}) in facility location or any other decisions, one first needs to provide an exact meaning of equity. Despite the importance of the subject,  there is no unique notion (definition) of equity that is generally accepted. Instead, there is a wide variety of notions of equity and fairness in the economics and decision theory literature that depend on the context. The primary concern for equity in resource allocation is treating entities fairly, such that everyone receives the same level of service and no one is at a disadvantage.  The allocated resource(s) can be a particular good, or bad, or a chance of good or bad. The entities can be a population, group of people at some location or belonging to some social classes or organizations, etc.  In general, most equity literature aims at equal distribution of benefits or disutilities between entities \citep{mostajabdaveh2019inequity}.

Although there is no single equity concept that we can use to design inequity-averse HCF location models, there are four key areas of equity research for health systems \citep{cardoso2015integrated, cardoso2016moving}. These are:

\begin{itemize}\itemsep0em
\item \textit{Equity of access}: Informally, accessibility is the relative ease by which patients can reach a healthcare facility from a given location \citep{hawthorne2013exploring, jin2015spatial, wang2012measurement}. Thus, patients should receive the care they need as close as possible to their place of residence or employment. 
In the case of emergency services, accessibility is the ability of a healthcare provider to reach the patients. Accessibility measures include both spatial and non-spatial factors \citep{wang2012measurement}. Spatial factors include the spatial separation between supply (e.g., surgical centers) and demand (e.g., patient population needing surgical care) and how they are connected in space. Thus it is a classical aspect in location analysis.  Non-spatial factors include demographic (e.g., age, gender, sex, etc.) and socioeconomic (e.g., income, poverty, female-headed households, etc.) variables, which also vary across geographical areas. 

\item \textit{Equity of utilization}: Utilization refers to the satisfied demand for different services. Ensuring equity of utilization means providing roughly equal service levels across services. Note that this diverges from the concept of ``\textit{deliver the cheapest service}'' often observed in location models that seek cost minimization.

\item  \textit{Socioeconomic equity}: Socioeconomic equity stipulates that the unsatisfied demand for population groups with lower income should not be greater than that of groups with higher income. Decision-makers may also want to ensure that unmet demand for lower-income or vulnerable population groups is sufficiently low. 

\item \textit{Geographical equity}: Geographical equity refers to the ability of the system to provide relatively equal levels of unsatisfied (satisfied) demand across geographical areas. Decision-makers often want to ensure that unsatisfied demand is not vastly different across geographical areas, or that some geographic areas do not lack healthcare service entirely.
\end{itemize}
\begin{rem}
Like any other metrics in any optimization problem, these four metrics can be targeted either by putting them in the objective function or in the constraints. For example, we might minimize the disparity in unsatisfied demand across income levels (socioeconomic equity) and/or add a constraint requiring the unsatisfied demand in any geographic area to be no more than a fixed maximum level (geographical equity).
\end{rem}

 Equity research in other application domains additionally considers \textit{social equity} and \textit{diversity}. The \textit{social equity} concept quantifies equity based on how any good received is proportional to the need \citep{levinson2010equity}. For example, the volume of the demand for a particular health service may differ among demand nodes in rural and urban areas.  If only a fraction of the demand can be satisfied, measures such as the proportion of the satisfied demand can be used to measure equity and service quality \citep{karsu2015inequity}. \textit{Diversity }is another concept that is indirectly related to equity. Suppose, for example, that we want to select a set of locations to open vaccination centers. The decision-maker may have concerns about diversity because they want certain population groups to have a certain degree of coverage or access to vaccination by the chosen location.   One way to achieve this is to use \textit{quotas}, i.e., ensuring that a certain proportion of the vaccination centers will be located to cover the groups of concern  \citep{karsu2015inequity}. This approach treats people with different characteristics differently, such that the selected locations are diversified enough in the sense that they cover diverse groups of concerns.

In contrast to most of the HCF literature (see Section~\ref{sec:EquityInHCF} and \cite{ahmadi2017survey}), inequity measurement has found explicit and extensive consideration in the economic and decision theory literature and a few discussions in the humanitarian logistics literature. The commonly accepted theme is that \textit{there is no one-size-fits-all solution to ensure equity, and customized methods are needed to measure and handle application-specific equity concerns}. Using transparent and explicit criteria that determine what is equitable and what is not is useful in ensuring that the decisions are acceptable, equitable,  and implementable in practice. 

There are also different operations research (OR) methods and metrics for incorporating equity in the decision process. The precise interpretation of each depends on both the structure of the problem at hand and the decision maker's understanding of equity \citep{karsu2015inequity}.   \cite{karsu2015inequity} give a comprehensive and deep survey on the use of equity concepts connected with optimization models. The authors distinguish between two types of inequity-averse optimization approaches, addressing ``\textit{equitability}''  and ``\textit{balance}.'' \textit{Equitability} is a concern whenever one deals with the allocation of goods across indistinguishable entities. In contrast,  \textit{balance} is a concern whenever goods are distributed across entities with different needs, claims, or preferences.   In the following subsections, we provide a high-level overview and analysis of the most common equity measures detailed in \cite{karsu2015inequity}.

\subsection{\textbf{The Rawlsian approach}}

The Rawlsian  approach \citep{rawls1999theory}  is one of the oldest, most common, and simplistic approaches used in OR to incorporate fairness or equity in optimization models. This approach represents equity preference by focusing on the worst-off entity, i.e., the minimum outcome level in a distribution. One can enforce a constraint that ensures that the minimum outcome is larger than a predefined level or seek to maximize the minimum outcome. For example, in the \textit{p}-center problem, we seek to locate \textit{p} facilities to minimize the maximum distance between any demand point and its nearest facility. As such, each demand point is assigned to the nearest facility, ensuring that each demand node is covered. Although the notion of coverage is well established in healthcare applications, in many HCF applications, we are often interested in the average distance that a patient has to travel to receive health service or the average distance that healthcare providers must travel to reach their patients \citep{daskin2005location}. Some studies extend the Rawlsian approach to a maximum lexicographic approach. That is, the welfare of the worst-off is first maximized subject to resource and other constraints, then the second worst-off, then the third worst-off, and so on \citep{kostreva2004equitable}.  As pointed out by \cite{karsu2015inequity}, the lexicographic maximin approach is a regularization of the Rawlsian maximin approach and is inequity averse.

\subsection{\textbf{Approaches based on inequity indices}}

Various studies that involve equity incorporate an inequity index into the model, which often assigns a scalar value to any given distribution showing the degree of inequity. Inequity indices are often used to assess the disparity in distribution, and so they are related to several mathematical concepts of dispersion and variance. Inequity indices respect the anonymity property \citep{chakravarty1999measuring} and often equal 0 when perfect equity occurs. \textit{Anonymity property} indicates that an inequity measure does not depend on the labeling of individuals. As pointed out by \cite{panzera2020measuring}, the anonymity property implies that an inequity measure is permutation invariant, which means that very different spatial patterns can give rise to the same measure. 
Suppose we have a set of $i \in I$ individuals (or demand points). Let $x_i$ donate the outcome value at node $i$. Below we use $x$ to briefly summarize the most commonly used inequity indices in the literature (adapted from \cite{karsu2015inequity}):
\begin{itemize}
\item \textit{The deviation from the mean}  ($\sum_{i \in I} (x_i-\bar{x})$, where $\bar x$ is the mean value).  This index measures the total deviation from the mean.  In some applications, the mean of the outcome distribution is often unknown, especially at the time when the decisions leading to the outcome $x$ are being made. Thus, the mean $\bar x$ is often approximated based on expert knowledge or derived endogenously in the model.  Some studies use the total absolute deviation from the mean (i.e., $\sum_{i \in I} \vert x_i- \bar{x} \vert$). Note that the mean absolute deviation (MAD) disregards how these deviations are distributed. Thus, MAD does not provide an incentive to minimize the gap among high values of the outcome (i.e., above or equal to the average) and among low values of the outcome (i.e., below or equal to the average). For example, deviations 0.25 and 0.25  are equivalent to deviations 0.5 and 0.  Other studies use the mean squared deviation, the maximum componentwise deviation from mean, or only the positive or negative deviation from the mean as a measure of inequity.  Mathematically, if the mean is (assumed) known, then using MAD in the objective function or a constraint may yield a linear optimization problem. For example, $\min \{ \sum_{i \in I} \vert x_i-\bar x \vert : x_i \in \calX\}$ is equivalent to $\min \{ \sum_{i \in I} z_i: z_i \geq x_i-\bar x, z_i \geq \bar x -x_i,  x_i \in \calX\}$, where $\calX$ is a set of (potentially) linear constraints on $x$.
\item \textit{The range, or difference between the minimum and maximum levels of outcomes }($\max_i x_i - \min_i x_i$).  Some studies also minimize this range normalized by the minimum outcome or enforce a constraint that ensures that $\frac{\min_{i}x_i}{\max_{i}x_i}  \geq \beta$, where $\beta$ is an equity or fairness parameter \cite{chang2006optimal}.  This index is used in many applications owing to its being simple and easy to understand.  However, as pointed out by \cite{karsu2015inequity}, by considering the two extremes (e.g., most and least deprived), this index is rather crude as it fails to distinguish allocations with the same values of extremes but different levels of other values. Thus, the range is sensitive to extreme values and ignores the interior of the distribution. Note that the normalized range may lead to a non-linear and complex formulation.

\item \textit{The variance or standard deviation.} A small variance means a low dispersion of the outcome. Both variance and standard deviation typically result in non-linear optimization problems.

\item \textit{The Gini Coefficient} ($ \sum_{i \in I} \sum_{j \in I} \vert x_i-x_j\vert/2 \vert I\vert \sum_{i \in I}x_i$). The Gini coefficient (and indices derived based on it) is one of the most widely used measures of income inequity in an economy that satisfies the Pigou–Dalton principle of transfers (PD), which states that any transfer from a poorer person to a richer person, other things remaining the same, should always lead to a less equitable allocation. Note that the Gini index is often a dimensionless quantity, and thus, it cannot often be incorporated in a natural way with other terms in a multi-criteria problem \citep{gutjahr2018equity}.  Moreover, this measure has the disadvantage of being highly non-linear, possibly making the resulting optimization problem extremely complex. The Gini index will always assume a value between 0 (indicating total equity) and 1 (indicating total inequity)
\item \textit{Sum of pairwise (absolute) differences} ($\sum_{i \in I} \sum_{j \in I}\vert x_i-x_j \vert$) and sum of squared deviations between all pairs. In contrast to MAD, the sum of pairwise absolute differences (SAD) does consider the spread of the outcome. Like MAD, the SAD term can be linearized (e.g., $\min \sum_{i \in I} \sum_{j \in I} \vert x_i-x_j \vert\equiv \min \{\sum_{i \in I} \sum_{j \in I}z_{i,j}: z_{i,j}\geq x_i -x_j, \ z_{i,j} \geq x_j-x_i, \forall i,j\}$.
\item \textit{The deviation from a predefined target}: If the predefined target is the best possible output, for example, then satisfying it indicates perfect equity. Thus the larger the deviation, the larger the inequity. Minimizing this deviation is often referred to as minimizing regret. Related measures include minimizing maximum regret, minimizing absolute regret, etc.

\end{itemize}

\begin{rem} Selecting one of the above (or other) indices implies a particular assumption on the decision maker's or optimizer's attitude to equity. 
\end{rem}

\subsection{\textbf{Approaches based on inequity-averse aggregation functions}}

\cite{karsu2015inequity} propose approaches based on inequity-averse aggregation functions, which use the aggregation function of the distribution vector in the model that would encourage equitable distributions. Unlike an inequity index, which only focuses on the inequity in a distribution, an inequity-averse aggregation function reflects concerns for both equity and efficiency. There are several ways to capture equity in this approach. For example, some studies use aggregation functions that have convenient mathematical properties such as convexity. 
\cite{marin2010extended} use  \textit{ordered median functions} as objective functions of  flexible discrete location problems. Ordered median functions are weighted total cost functions, in which the weights are rank-dependent. Ordered median functions with appropriate weights are inequity-averse in the sense that they are strictly concave.  We refer to \cite{karsu2015inequity} for more details on inequity-averse aggregation functions.

\section{Equity versus Uncertainty}\label{Sec3:EquityVsUncer}

\noindent Uncertainty is intrinsic to many HCF location problems since various key input parameters such as demand, costs, and travel times are often unpredictable. While inequity-averse optimization in a deterministic context is conceptually relatively simple, though often computationally nontrivial and demanding, location decisions under uncertainty represented by suitable stochastic models introduce additional challenges for the following primary reasons.  First,  because of uncertainty, the value estimates are often not perfectly accurate, whereby the ex-post realized values of the impact (e.g., access to care, percentage of satisfied demand across geographical locations) and alternative decisions rarely coincide with their ex-ante estimated values.

Second, in most real-world applications such as HCF location, it is unlikely that we can accurately infer the actual distributions of random parameters and thus quantify the impacts of decisions on equity, especially with limited data or no information during the planning stage. Even when historical data is available, the quality of such data may not be sufficient to estimate the distribution of uncertain factors accurately, and  future uncertainty is often not distributed as the past. Various studies show that different distributions can typically explain raw data of uncertain parameters, indicating distributional ambiguity (i.e., uncertainty in distribution type \citep{esfahani2018data, vilkkumaa2021causes}).

Third, incorporating equity measures and addressing both uncertainty and distributional ambiguity may increase the overall complexity of HCF location problems. However, ignoring uncertainty and equity--uncertainty interaction may lead to devastating costs and health outcomes. Adverse outcomes associated with poor HCF location decisions include increased costs, disparities in service, and increased illness or death. For example, a hard-to-access healthcare facility is likely to be associated with increased morbidity (disease) and mortality (death).

In this section, we demonstrate through a simple example that uncertainty and equity interact in ways that should not be ignored---the decision one should make in the presence of both considerations is often different from the decision under either consideration in isolation. As detailed \cite{ahmadi2017survey} and \cite{daskin2005location}, and in Section~\ref{sec:EquityInHCF},  most of the existing HCF location models are extensions of basic discrete location problems (in which facilities can be established only at candidate locations), including \textit{p}-median, \textit{p}-center, covering-based, and fixed charge models. For brevity and illustrative purposes, herein, we next analyze extensions of the \textit{p}-median and \textit{p}-center using some of the linear equity metrics discussed in the previous section. Specifically, we define different equity objectives based on the sum or maximum of pairwise (absolute) differences in distances or demand-weighted distances.

Table~\ref{table:notation} summarizes the general notation we use in all formulations. The decision variables listed in the table are first-stage decision variables; we will introduce the second-stage variables shortly, when we discuss stochastic models. We assume that all demand nodes are also potential facility locations, denoted by $I$, but this assumption is straightforward to relax if necessary. Using the notation in the table~\ref{table:notation}, we define the following common feasible region among all formulations:

\begin{table}[t!]  
\small
\center
   \renewcommand{\arraystretch}{0.67}
  \caption{General Notation.} 
\begin{tabular}{ll}
\hline
\multicolumn{2}{l}{\textbf{Parameters and sets}} \\
$p$ & number of facilities \\
$I$& number, or set, of locations \\
$d_{i,j}$ & distance/travel time between any pair of nodes $i$ and $j$ \\
$w_{i}$& demand at node  $i$ \\
 \multicolumn{2}{l}{\textbf{First-stage decision variables } } \\
$x_{j}$ &   $\left\{\begin{array}{ll}
1, & \mbox{if a facility is open at  candidate  location } j, \\
0, & \mbox{otherwise.}
\end{array}\right.$ \\
$y_{i,j}$ &   $\left\{\begin{array}{ll}
1, & \mbox{if demand point \textit{i} is assigned to a facility at candidate location } j, \\
0, & \mbox{otherwise.}
\end{array}\right.$ \\
\hline
\end{tabular}\label{table:notation}
\end{table} 

 \begin{align*}
\mathcal{X}&=\left\{ x, y : \begin{array}{l} (\text{C1}) \ \ \sum_{j \in I} x_j=p \\ 
  (\text{C2}) \ \ \sum_{j \in I} y_{i,j}=1, \forall i \in I  \\ 
  (\text{C3}) \ \  y_{i,j} \leq x_{j},   i \in I, j \in I \\
 (\text{C4}) \ \  y_{i,j} \in \{0,1\}, \ x_j  \in \{0,1\}, \forall i \in I, j \in I 
    \end{array}\right\}  
\end{align*} 

Constraint (C1) specifies the total number of facilities to be established. Constraints  (\text{C2})  ensure that each demand point is assigned to only one facility, and constraints (\text{C3})  limit assignments to open facilities.  Constraints (\text{C4}) are integrality constraints. 

Table~\ref{Table:Models} presents several deterministic formulations for locating \textit{p} facilities, including both classical facility location problems such as the $p$-median and $p$-center, as well as equity-based formulations. We do not claim that the equity-based approaches listed here are the best models for considering equity. Rather, we chose these measures because they are used frequently and because they are useful for illustrating the interaction between uncertainty and equity.

In the next section, we compare the optimal solutions obtained using the deterministic formulations in Table~\ref{Table:Models} and their stochastic programming (SP) counterparts. In the SP, travel time and demand are modeled as random variables that follow fully known probability distributions. The objective is to minimize the expectation of the objective function, where the expectation is taken with respect to an assumed known distribution.  We approximate solutions to the SP models using their sample average approximation (SAA).  That is, we generate a sample of $N$ scenarios (each scenario consists of a vector of realizations of demand and travel time which are drawn independently from the distributions corresponding to each node and pair of nodes, respectively), and then optimize the sample average of the objective. (The technical details of SAA  are out of the scope of this chapter, and we refer the reader to \cite{kim2015guide, kleywegt2002sample, mak1999monte, shapiro2021lectures}, and references therein, for a thorough discussion.). For example, the SAA of the \textit{p}-center model is as follows:
\begin{subequations}
\begin{align}
\min  \left\{\sum_{n=1}^N \frac{1}{N} z^n :  (x,y) \in \calX, \ \ z^n \geq \sum_{j \in I}d_{i,j}^n y_{i,j}, \ \ \forall i \in I, \ n \in N \right\}
\end{align}
\end{subequations}
Here, the $d^n_{i,j}$ values are the travel times (distances) under the $n$th sample, and $z^n$ is the optimal objective function value of the (deterministic) $p$-center problem under the $n$th sample.

\begin{table}[t!]
 \center 
 \footnotesize
 \renewcommand{\arraystretch}{0.6}
 \caption{Optimization Models}
\begin{tabular}{llll}
\hline
\textbf{Model Name} 	&	\textbf{Formulation }	\\
\hline 
\textit{p}-median &  $\min  \Big \{\sum \limits_{i \in I} \sum \limits_{j \in I} w_i d_{i,j}y_{i,j}: (x,y) \in \calX  \Big \}$ \\
\textit{p}-center & $\min  \Big \{  \max \limits_{i \in I} \sum \limits_{j \in I} d_{i,j}y_{i,j}:  (x,y) \in \calX \Big  \}$
 \\
Total distance &  $\min  \Big \{\sum \limits_{i \in I} \sum \limits_{j \in I} d_{i,j}y_{i,j}: (x,y) \in \calX  \Big \}$\\
Equity 1 & $\min   \Big \{\sum \limits_{i \in I} \sum \limits_{j \neq i \in I} \vert z_i-z_j \vert :  z_i =\sum  \limits_{j \in I} d_{i,j}y_{i,j}, \ \forall i \in I  \Big \}$ \\
Equity 2 &  $\min \Big\{ \max \limits_{i,j} \vert z_i-z_j\vert : (x,y) \in \calX, \  z_i =\sum  \limits_{j \in I}  d_{i,j}y_{i,j}, \ \forall i \in I  \Big \} $\\
Equity 3& $\min \Big\{  \sum \limits_{i \in I} \sum \limits_{j \neq i \in I}\vert z_i-z_j\vert :  (x,y) \in \calX, \ z_i =\sum  \limits_{j \in I} w_i d_{i,j}y_{i,j}, \   i \in I  \Big \} $\\
Equity 4 & $\min\Big\{  \max \limits_{i,j} \vert z_i-z_j\vert :  (x,y) \in \calX, \ z_i =\sum  \limits_{j \in I} w_i d_{i,j}y_{i,j}, \   i \in I  \Big \}$ \\
Equity 5 & $\min\Big\{ \max \limits_{i \in I} \sum  \limits_{j \in I} \vert z_i-z_j\vert :  (x,y) \in \calX,  \ z_i =\sum  \limits_{j \in I}  d_{i,j}y_{i,j} , \  \forall i \in I  \Big \}$ \\
Equity 6 & $\min\Big\{  \max \limits_{i \in I} \sum \limits_{j \in I} \vert z_i-z_j\vert :  (x,y) \in \calX, \  z_i =\sum  \limits_{j \in I}  w_id_{i,j}y_{i,j},     \  \forall i \in I  \Big \}$ \\
Equity 7 & $ \min\Big\{  \sum \limits_{i \in I} \max \limits_{j \in I} \vert z_i-z_j\vert  :   (x,y) \in \calX, \ z_i =\sum  \limits_{j \in I}  d_{i,j}y_{i,j},     \  \forall i \in I  \Big \}$ \\
Equity 8 & $ \min\Big\{  \sum \limits_{i \in I} \max \limits_{j \in I} \vert z_i-z_j\vert  :   (x,y) \in \calX, \ z_i =\sum  \limits_{j \in I}  w_id_{i,j}y_{i,j},     \  \forall i \in I  \Big \}$ \\
\hline 																		
\end{tabular} 
\label{Table:Models}
\end{table}

\subsection{Example: Where to locate a new hospital in Lehigh County?}

\noindent   In this subsection, we consider locating a single hospital (i.e., $p=1$) in a service region based on Lehigh County, 
which is located in the Lehigh Valley region of the U.S.\ state of Pennsylvania. We consider a dataset consisting of 21 nodes: the 20 largest communities in Lehigh County according to the 2010 census, plus the city of Easton.\footnote{Easton is not in Lehigh County; we included it because it is the third largest city in the Lehigh Valley.}  We calculated the distance and travel time between each pair of nodes using the Google API. 
We use these travel times as the $d_{i,j}$ values for the deterministic models. For the stochastic models we set both the average travel time ($\mu^{\tiny d}_{i,j}$) and the standard deviation of travel times ($\sigma^{\tiny d}_{i,j}$) between each pair of nodes $(i,j)$ equal to the calculated travel time, for all $(i,j)$. 

We use the population estimate in each county based on the 2010 U.S.\ census (see Table~\ref{table:LehighInst}) to construct the following demand structure.  We use the population percentage (weight) at each node to generate the mean (average)  demand at each node $i \in I$ as $\mu^{\tiny w}_i=\texttt{population\%} \times 1000$ (i.e., total demand of 1000). To a certain extent, this structure reflects what may be observed in real life, i.e., locations with more population typically create greater demand. We set the standard deviation as $\sigma^{\tiny w}_{i}=0.5\mu^{\tiny w}_{i}$, for all $i \in I$.

We generate the following two sets of $N$ data samples for the parameters $w$ and $d$.
    \begin{itemize}
        \item Set 1: $w_i \sim$ lognormal (LogN) with mean $\mu^{\tiny w}_i$ and standard deviation $\sigma^{\tiny w}_i$, and  $d_{i,j} \sim U[\mu^{\tiny d }_{i,j}-\Delta,\mu^{\tiny d }_{i,j}+\Delta]$, where $\Delta=10$ minutes.
        \item Set 2: $w_i \sim$ LogN with mean $\mu^{\tiny w}_i$ and standard deviation $\sigma^{\tiny w}_i$, and $d_{i,j} \sim $ LogN with mean $\mu^{\tiny d}_{i,j}$ and standard deviation $\sigma^{\tiny d}_{i,j}$.
    \end{itemize}  
We solve the SAA counterparts of the deterministic  models in Table~\ref{Table:Models} with the generated data samples.  In addition, we solve the deterministic formulations with one scenario ($N=1$).

Tables \ref{Table:Set1} and \ref{Table:Set3} present the optimal solutions under Sets 1 and 2. We make the following observation from these Tables. First, the optimal location can change when we consider uncertainty \textit{or} equity, which is expected and not new.   Second, incorporating both uncertainty and equity result in a solution that is often different from incorporating either one. For example, the deterministic and SAA solutions of the \textit{p}-median problem respectively locate the hospital at Catasauqua and Fountain Hill under Set 1. In contrast, the deterministic and SAA solutions of the Equity 3 problem, which minimizes the sum of the absolute deviations in demand-weighted travel time, respectively locate the hospital at Allentown and Dorneyville.  Third, different measures of inequity aversion under uncertainty can result in different optimal solutions.  Fourth, the SAA solutions may be different under different distributions, motivating the need for distribution-free models. For example, the SAA solution for Equity 2 locates the hospital at  Catasauqua and Cetronia under Set 1 and Set 2, respectively.

\begin{table}[t!]
 \center 
 \footnotesize
   \renewcommand{\arraystretch}{0.6}
  \caption{Lehigh County nodes and their population based on the 2010 census of Lehigh County.}
\begin{tabular}{lllc}
\hline
\textbf{City/Town/etc.} 	&	Pop	&	Pop\%	&	Avg. Demand	\\
\hline 
Allentown	&	118,032	&	40.9\%	&	409	\\
Bethlehem	&	74,982	&	26.0\%	&	260	\\
Emmaus	&	11,211	&	3.9\%	&	39	\\
Ancient Oaks	&	6,661	&	2.3\%	&	23	\\
Catasauqua	&	6,436	&	2.2\%	&	22	\\
Wescosville	&	5,872	&	2.0\%	&	20	\\
Fountain Hill	&	4,597	&	1.6\%	&	16	\\
Dorneyville	&	4,406	&	1.5\%	&	15	\\
Slatington	&	4,232	&	1.5\%	&	15	\\
Breinigsville	&	4,138	&	1.4\%	&	14	\\
Coplay	&	3,192	&	1.1\%	&	11	\\
Macungie	&	3,074	&	1.1\%	&	11	\\
Schnecksville	&	2,935	&	1.0\%	&	10	\\
Coopersburg	&	2,386	&	0.8\%	&	8	\\
Alburtis	&	2,361	&	0.8\%	&	8	\\
Cetronia	&	2,115	&	0.7\%	&	7	\\
Trexlertown	&	1,988	&	0.7\%	&	7	\\
Laurys Station	&	1,243	&	0.4\%	&	4	\\
New Tripoli	&	898	&	0.3\%	&	3	\\
Slatedale	&	751	&	0.3\%	&	3	\\
Easton	&	26,800	&	9.3\%	&	93	\\
\hline 
Total 	&	288,310	&		&		\\
\hline 																		
\end{tabular} 
\label{table:LehighInst}
\end{table}

\begin{table}[t!]
 \center 
 \footnotesize
 \renewcommand{\arraystretch}{0.6}
  \caption{Optimal locations yielded by each model under Set 1. Notation: DET is the deterministic model solved with 1 scenario, and SAA is the SAA counterpart of each  solved with $N=50.$}
\begin{tabular}{llll}
\hline
Model	&	DET	&SAA 	\\
\hline 
p-median 	&	Catasauqua (5)	&	Fountain Hill (7)	\\
p-center  	&	Dorneyville (8)	&	Catasauqua (5)	\\
Total Distance	&	Dorneyville (8)	&	Cetronia (16)	\\
Equity 1	&	Coplay (11)	&	Cetronia (16)	\\
Equity 2  	&	Allentown (1)	&	Catasauqua (5)	\\
Equity 3 	&	Allentown (1)	&	Dorneyville (8)	\\
Equity 4	&	Catasauqua (5)	&	Dorneyville (8)	\\
Equity 5	&	Allentown (1)	&	Catasauqua (5)	\\
Equity 6	&	Catasauqua (5)	&	Dorneyville (8)	\\
Equity 7	&	Catasauqua (5)	&	Catasauqua (5)	\\
Equity 8	&	Catasauqua (5)	&	Fountain Hill (7)	\\
\hline 																		
\end{tabular} 
\label{Table:Set1}
\end{table}


\begin{table}[t!]
 \center 
 \footnotesize
 \renewcommand{\arraystretch}{0.6}
  \caption{Optimal locations yielded by each model under Set 2.}
\begin{tabular}{llll}
\hline
Model	&	DET	&SAA 	\\
\hline 
p-median 	&	Cetronia (16)	&	Allentown (1)	\\
p-center  	&	Catasauqua (5)	&	Cetronia (16)	\\
Total Distance	&	Wescosville (6)	&	Dorneyville (8)	\\
Equity 1	&	Catasauqua (5)	&	Cetronia (16)	\\
Equity 2  	&	Catasauqua (5)	&	Cetronia (16)	\\
Equity 3 	&	Cetronia (16)	&	Allentown (1)	\\
Equity 4	&	Cetronia (16)	&	Allentown (1)	\\
Equity 5	&	Catasauqua (5)	&	Cetronia (16)	\\
Equity 6	&	Cetronia (16)	&	Allentown (1)	\\
Equity 7	&	Catasauqua (5)	&	Catasauqua (5)	\\
Equity 8	&	Cetronia (16)	&	Allentown (1)	\\
\hline 																		
\end{tabular} 
\label{Table:Set3}
\end{table}
\section{Is the Stochastic HCFL Literature Inequity-averse?}\label{sec:EquityInHCF}

In this section, we provide a high-level analysis of recent stochastic approaches for HCF location, focusing on studies proposing inequity-averse approaches published between 2004 and 2017. By inequity-averse, we mean any approach that considers one or more of the considerations mentioned above or other equity-related objectives or constraints. Our goal is to bring attention to a fundamental and timely question: \textit{Is the Stochastic HCF Location Literature Inequity-averse?}. We next analyze the limited literature considering equity and uncertainty, highlighting existing equity-related objectives or constraints and the challenges of incorporating these.

For a comprehensive survey on the HCF location--allocation literature, we refer to \citep{rahman2000use, daskin2005location,li2011covering, gunecs2019location,cisse2017or, gutierrez2013home, grieco2020operational, ahmadi2017survey}. The recent survey by Ahmadi-Javid et al.\ \cite{ahmadi2017survey} provides a thorough classification of HCF location problems, models, and solution methods in the last decade, identifying gaps and possible future directions. They first provide a framework to classify different types of non-emergency and emergency HCFs. Then, they analyze the literature on HCF location problems along ten descriptive dimensions (e.g.,  uncertainty, single or multi-period settings, etc.). Next, we dive deeper into this literature, highlighting those considering equity and/or uncertainty. Note that mathematical differences between optimization models for each type (and sub-type) of HCFs are due to, for example, the nature of the service provided (and thus different objectives, constraints, and random factors), nature of the operation (mobile vs. static,e.g., mobile primary care clinic vs. a primary clinic in a hospital), decision-maker perspective, case study, etc.


\subsection{\textbf{Non-emergency HCF location}}

\noindent Non-emergency medical (health) services include medical treatment, observation, prevention, testing, and other healthcare services provided to patients whose conditions are not urgent or considered an emergency. Non-emergency facilities include primary care facilities (hospitals, outpatient and primary care clinics, etc.), blood banks, specialized centers (e.g., organ transplant centers), medical laboratories, mobile healthcare units, drugstores, etc. Next, we review some of the literature on major non-emergency HCFs.

\subsubsection{\textbf{Primary care facilities}}

\noindent Primary care facilities (PCF)  is a class of HCF that provide primary care to the public, including early diagnosis and first-contact care.  Most PCFs (e.g., hospitals, clinics) are open for 24 hours, and patients often tend to visit the nearest one. The optimal location of these facilities has received significant attention from the operations research community. Basic location models used include set covering, maximal covering, $p$-median, and fixed-charge.  Within this literature, most papers proposed deterministic models. \cite{mitropoulos2006biobjective} proposed a bi-objective mathematical programming model for locating hospitals and primary healthcare centers. The two objectives in this model are: (1) minimization of the distance between patients and facilities, (2) equitable distribution of the facilities among citizens. \cite{gunecs2014matching} also considered minimizing the maximum travel distance in designing a primary care facility network. Other (deterministic) studies minimize deviations from a standard distance \citep{smith2013bicriteria}.

\cite{oliveira2006modelling} proposed two mathematical programming location--allocation models to redistribute hospital supply using different objective functions and assumptions about the utilization behavior of patients. The first model optimizes equity by minimizing variations between predicted and normative utilization (according to need) by small area. The second model optimizes equity by minimizing utilization flows between small areas and hospitals and a utilization flows target (defined flows of patients using closest and central hospitals).

\cite{rahmaniani2014variable}  proposed a multiobjective two-stage stochastic nonlinear integer programming model for the location--allocation of hospitals. This model's objective is to minimize total travel, operating costs, and congestion costs. Uncertainty considered includes the fixed cost of opening a facility, travel time (distance) between nodes, the capacity of facilities, and demand. They used the expected demand weighted travel time as a measure of random accessibility. Due to the challenges of solving the model exactly, \cite{rahmaniani2014variable} proposed a heuristic solution algorithm based on variable neighborhood search (VNS).

\cite{mestre2015location} propose two location--allocation models for handling demand uncertainty in the strategic design of a hospital network. In addition to operational objectives, both models aim to maximize access by minimizing the expected travel time to reach hospital services weighted by demand. \cite{mccoy2014using} propose equitable allocation strategies for motorcycle trips facilitating access to healthcare in rural areas.  \cite{ares2016column} used a coverage score to model equity in terms of access to healthcare among the different populations. \cite{beheshtifar2015multiobjective} proposed a new definition for equity by minimizing the variability of access distance to a healthcare clinic, where variability was measured in terms of the standard deviation of distances from the place of demand points to the related open site.  \cite{shishebori2015robust} proposed a mixed-integer linear programming model for a robust and reliable medical service center location network design problem,  which simultaneously takes uncertainty in demand, transfer cost, and system disruptions into account. However, the model did not include any explicit equity measure.

\subsubsection{\textbf{Blood Banks}}

\noindent A blood bank is a free-standing HCF or part of a larger HCF (e.g., hospital) that collects blood samples from donors, then stores and prepares them  for transfusion to recipients. Timely access to different blood banks (e.g., blood transfusion providers, blood centers, mobile blood banks units) is crucial in health systems. One of the main challenges to blood bank location is that human blood is scarce, perishable, and often in high demand (e.g., for cancer patients, surgical patients, patients presenting to emergency rooms, etc.).  Furthermore, both demand and supply of blood are stochastic and subject to various disruptions.

\cite{jabbarzadeh2014dynamic} presents a robust location--allocation model for dynamic supply chain network design for the supply of blood in disasters. Uncertainty considered includes the capacity of a temporary blood facility, capacity of a permanent blood facility, and maximum blood supply of each donor group.  The objective is to minimize the total cost of the network (the combined costs of locating blood facilities, transportation, and holding), while ensuring that the network is robust to major disasters. \cite{fahimnia2017supply} also proposed a stochastic bi-objective supply chain design model for the efficient (cost-minimizing) and effective (delivery time minimizing) blood supply in disasters. Uncertainty considered includes the` cost of moving a mobile blood facility, unit operational cost at each mobile blood facility, local blood center, and regional blood center, unit transportation cost between blood facilities, and blood demand at each hospital.  Although both studies considered uncertainty, they did not incorporate any equity objective or constraints.

\subsubsection{\textbf{Organ transplant centers}}

\noindent Organ transplant centers (OTC) are the main components of organ transplantation programs in in most healthcare systems. The demand for organs is a major random factor that is often larger than the supply, which is also random. As a result, organ transplants suffer from long waiting lists. The time between the request for an organ and transplantation, transportation time of organs from donors' locations (e.g., hospital) to OTC, and transportation time of recipients to OTC are vital in the process of organ donation/transplantation and subject to a high degree of uncertainty. \cite{zahiri2014robust} present a robust probabilistic programming approach to multi-period location--allocation of OTCs. The objective is to minimize the weighted total costs, including fixed opening costs, costs of the removal process, transportation costs, unsatisfied demands, and the saving costs associated with integrating hospitals and  organ transplant centers  at the same location. They demonstrate that solutions derived using their model are robust when taking into account uncertain conditions in the form of small yearly demand changes. \cite{zahiri2014multi} propose a multi-period location--allocation bi-objective mathematical programming model for designing an organ transplant transportation network under uncertainty. The model minimizes total cost and time, including waiting time in the queue for the transplant operation while considering organs' priorities. Uncertainty considered includes inter-arrival times of organs entering the transplant centers, arrivals of patients to the transplant centers, among others. These studies did not consider any equity objective or constraints. 

\subsubsection{\textbf{Detection and prevention centers}}

\noindent Detection and prevention centers are HCFs that provide healthcare services defined based on local or national detection and prevention programs \cite{ahmadi2017survey}. The common aim of these HCFs is to reduce the likelihood and severity of potentially life-threatening illnesses by protection and early detection. Therefore, the level of participation in preventive healthcare programs is crucial in their effectiveness and efficiency, so most studies considered participation maximization objectives. Other objectives include minimizing travel distance or time to increase accessibility and thus participation.

\cite{zhang2009incorporating} argue that the number of facilities to be established and the location of each facility are the main determinants of the configuration of an HCF network. They use the total travel time, waiting time, and service time required for receiving the preventive service as a proxy for accessibility of a healthcare facility and assume that each patient would seek the facility's services with a minimum expected total of these metrics.  To capture the congestion level, they modeled each facility as an M/M/1 queue. They show that the expected number of participants from each population zone decreases with the expected total time.  The objective is to find the optimal set of locations to maximize the total number of participants who request (and receive) service from the facilities, where facilities cannot be operated unless they achieve minimum workload requirements. They included a constraint in the model that ensures equity among the people living in a population zone in terms of access to preventive health services. \cite{zhang2009incorporating} is highly nonlinear and challenging. Therefore, they provide a heuristic solution framework for this problem.

\cite{zhang2010bilevel} proposed a bilevel model for preventive healthcare facility network design with congestion. They formulate the lower-level problem, which determines the allocation of clients to facilities, as a variational inequality, while the upper level is a facility location and capacity allocation problem.  Major random factors considered include the demand rate at each zone, travel times, and service times. The objective is to maximize participation by locating an undetermined number of facilities at the population zones over the network and allocating a given number of servers to open facilities. They incorporate congestion at the facilities in the model and assume that clients patronize the facility with the minimum expected total time (travel time to the facility and the expected time spent at the facility for possibly waiting and receiving the service). Thus, they considered the total time needed to receive preventive services at a facility as a proxy for its accessibility and did not include equity objectives or constraints.

\cite{vidyarthi2015impact} analyzed the impact of system-optimal (i.e., directed) choice on the design of the preventive healthcare facility network under congestion. The problem is set up as a network of spatially distributed M/G/1 queues and formulated as a nonlinear mixed-integer programming model. The model simultaneously determines the location and size of the facilities and the allocation of clients to these facilities to minimize the weighted sum of the total travel time and the congestion associated with waiting and service delay at the facilities. They linearize the model and present a cutting plane-based exact ($\epsilon$-optimal) solution approach to solve the reformulation. While the model in \cite{vidyarthi2015impact} can accommodate randomness in service and arrival rates, it does not capture the seasonality of the demand. Moreover,  \cite{vidyarthi2015impact}  did not include any equity objectives or constraints.

\cite{aboolian2016maximal} focus on designing facility networks in the public sector to maximize the number of people benefiting from their services. They propose an analytical framework for the maximal accessibility network design problem that involves determining the optimal number, locations, and capacities of a network of public sector facilities. They assume that the time spent for receiving the service from a facility is a good proxy for its accessibility. They assume that each node generates a stream of Poisson demands with a homogeneous rate and exponantial service time at each facility. \cite{aboolian2016maximal} did not consider equity objectives or constraints.


\subsubsection{Medical laboratories}

\noindent A medical laboratory is a HCF where tests are carried out on clinical specimens from the patient to aid in diagnosis, treatment, and disease prevention. Although medical laboratories are critical for public health, there is a lack of research on the location of these HCFs.  Notably, \cite{saveh2012setting} propose a $p$-median-based model for designing a network of tuberculosis testing laboratories to reduce transportation times and thereby decrease overall test turnaround time. They use the travel time from any region to any laboratory as a measure of equity. Accordingly, they included a constraint that allows decision-makers to specify an upper bound for origin--destination transportation time. Their results suggest that the optimal locations and capacities are not sensitive to this additional equity constraint. Modeling uncertainty was suggested as a future direction.

\subsubsection{\textbf{Long-term care centers}}
  
\noindent Long-term (nursing) care is an HCF that provides rehabilitative, restorative, and ongoing nursing care to patients or residents who need assistance with their health or daily living activities. The location of long-term care facilities is crucial to provide the best and most equitable possible services to aged people who represent the major demand group needing social and medical services. Given that this type of HCF provides medical care and social services to inpatients, the simultaneous determination of location, optimal capacity levels (e.g., number of beds), inventory levels, and locations are essential.
  
\cite{cardoso2015integrated} developed a fixed charge facility location-based model for planning a long-term care network, which considers demand uncertainty, multiple services, and various forms of equity (access, utilization, socioeconomic, and geographical equities) constraints.   \cite{cardoso2016moving} consider equity of access, geographical equity, and socioeconomic equity in long-term care (LTC) and network design decisions. They use minimization of total travel time for individuals accessing institutional LTC services to ensure equity of access, minimization of unmet need in the geographical area with the highest level of unmet need to ensure geographical equity, and minimization of unmet need for the lower-income groups to ensure socioeconomic equity in their model.

\subsubsection{Home healthcare centers, rehabilitation centers, doctors' offices, and drugstores}

\noindent Home healthcare (HHC) services provide healthcare services to people in their homes. HHC emerged in the early 50s to reduce the cost of care and health systems and improve patients' quality of life.  The home service industry, especially home healthcare, has been rapidly growing worldwide due to emergent changes in family structures, work obligations, aging populations, and the outspread of chronic and infectious diseases. Hence, the operations research community investigated different challenges of the HHC services, such as routing, appointment scheduling, staffing, and various resource allocation issues. However, as pointed out by \cite{ahmadi2017survey}, no studies exist for HHC center locations. It follows that there are no studies considering equity in HHC center locations.

Similarly, the location of rehabilitation HCF  (i.e., HCF devoted to the physical rehabilitation of patients with, e.g., neurological, orthopedic other medical conditions), doctors' offices, and drugstores are not studied as other types of HCFs. An analysis of equity concerns related to, for example, the distribution of these HCFs and access to them under uncertainty is an important future research area.

\subsection{\textbf{Emergency HCF location}}

Life-threatening emergencies, such as a severe injury, stroke, or heart attack, require the services of emergency HCFs. In addition, emergency HCFs provide service for patients with an injury or illness that does not appear to be life-threatening, but the treatment of such patients cannot wait until the next day or for a primary care doctor to see them. \noindent\cite{ahmadi2017survey} classify emergency HCFs according to whether they perform under permanent or temporary emergencies. \textit{Permanent} emergency HCFs provide service regularly, including emergency off-site public access devices, trauma centers, and ambulance stations. In contrast,  \textit{temporary} emergency HCFs are constructed to respond to unexpected health (e.g., infectious disease outbreak) and other situations (e.g., disaster). Next, we analyze the literature on emergency and trauma centers, ambulance stations, and temporary medical centers.

\subsubsection{Emergency and trauma centers}

\noindent Emergency departments or emergency centers are permanent emergency facilities that provide medical and surgical care to both patients arriving in need of immediate care or walk-ins (unscheduled patients). They can be part of a hospital or free-standing. Locating these facilities has not received as much attention as other emergency facilities such as ambulance stations.  Therefore, there is a need for inequity-averse stochastic models for the location of emergency centers, particularly those associated with hospitals and clinics or ambulance stations.  \cite{silva2008locating} is one of the early studies that presented a priority queuing-based covering location problem for locating emergency services considering different service priority levels. They implemented a greedy randomized adaptive search procedure to solve randomly-generated problem instances. \cite{silva2008locating} did not focus on equity.

Trauma centers are hospitals that provide specialized medical and nursing care to patients suffering from major traumatic injuries (e.g., falls, motor vehicle collisions and accidents, gunshot wounds, etc.). Patients are typically transported to trauma centers via helicopters or ambulances. To account for air transportation, some studies consider a joint location problem of trauma centers and helicopters under some budget constraints. Most existing studies for locating trauma centers employ maximal coverage location or fixed charge location models. \cite{ahmadi2017survey} propose a maximal backup coverage model  (BACOP) for the joint location problem of trauma centers and helicopters with budget constraints. Although trauma centers are rife with uncertainty (especially in demand), none of the papers reported in \cite{ahmadi2017survey} has incorporated uncertainty. In addition, equity and equity--uncertainty interaction has not been considered.

\subsubsection{Ambulance stations}

\noindent Emergency Medical Services, more commonly known as EMS, is a system that provides out-of-hospital acute medical care and transfers patients to emergency centers/departments within or outside hospitals and trauma centers for definitive  care. EMS aims to reduce the elapsed time to respond to an emergency. Ambulance stations and ambulances are major resources to achieve this aim. Ambulance stations are responsible for dispatching ambulances that provide EMSs on the scene or during transport to an emergency center/department, a trauma center, etc. The ambulance then returns to a predetermined station to await another call. For an efficient and timely response, ambulance stations must be located at appropriate points to provide adequate coverage and minimize the response time. The OR community has paid significant attention to the location of ambulance stations, the deployment (location, relocation, fleet sizing) of ambulances in the stations, and the dispatch of ambulances to the demand points or emergency sites.

Most of the existing models for this type of HCF are extensions of the basic maximum or set covering location, maximum expected coverage location \cite{Daskin1982,Daskin1983}, and $p$-center location models. Note that by focusing on maximizing the demand that can be covered, traditional covering models favor locating ambulances in more densely populated areas, resulting in longer response times for patients in more rural areas. That is, traditional covering models may lead to solutions in which the coverage pattern is quite good for those nodes counted as covered but extremely poor for those not covered, highlighting the need for equity-based models.

Most existing models for EMS are stochastic due to the stochastic nature of EMS operations. Random factors include, but are not limited to, the busy fraction of ambulances, demand or service requests, and travel time \citep{beraldi2004designing,beraldi2009probabilistic, gendreau2006maximal,ingolfsson2008optimal,rajagopalan2008multiperiod, mclay2009maximum,rajagopalan2009minimum,sorensen2010integrating,
noyan2010alternate, rajagopalan2011ambulance, naoum2013stochastic, zhang2015novel, yoon2021stochastic}.

\cite{noyan2010alternate} considers an EMS system design problem with stochastic demand. They proposed a capacitated fixed charge facility-like location model to locate the emergency response facilities and vehicles to ensure target levels of coverage, which are quantified using risk measures on random unmet demand. The model considers target service levels for each demand site and also for the entire service area.  \cite{noyan2010alternate} argues that considering the individual target service levels may be regarded as an alternative approach to model the coverage equity. The objective is to minimize the sum of the variable transportation costs, the fixed setup cost for opening the facilities, and the total cost of purchasing and maintaining vehicles. To present risk preferences,  they develop two types of stochastic optimization models involving alternate risk measures: integrated chance constraints (ICCs) and ICCs with a stochastic dominance constraint.

\cite{chanta2011minimum} proposed a minimum\textit{ p-envy} facility location model, aiming to find optimal locations for EMS facilities to balance customers' perceptions of equity in receiving service. Specifically, to deal with the issue of equity, they assigned an envy function (a function of the distance from a demand zone to its closest EMS station and the distance from a demand zone to its backup EMS stations weighted by priority of the serving stations and weighted by the proportion of demand) to each pair of demand nodes, for each level of priority.   This value indicates the dissatisfaction level of a demand node with its serving station in comparison with other demand nodes that have the same level of priority.

To address the issue of fairness in semi-rural/semi-urban communities, \cite{chanta2014improving}  propose a bi-objective covering location model for locating EMS ambulances at preexisting rescue stations that balances efficiency (i.e., maximizing expected coverage) and equity. Specifically, they propose the following alternative objective functions for improving fairness in rural areas: minimize the maximum distance between uncovered demand zones and their closest opened station, minimize the number of uncovered rural demand zones, and minimize the number of uncovered demand zones. \cite{chanta2014improving} use the $\epsilon$-constraint method to solve their multi-objective model.

\cite{khodaparasti2016balancing} studied balancing efficiency and equity in a location--allocation EMS model under uncertainty using data envelopment analysis.  \cite{enayati2019identifying} proposed a multicriteria optimization approach to study the trade-offs in equity and efficiency for simultaneously optimizing location and multipriority dispatch of ambulances. 


\subsubsection{Temporary medical centers}

\noindent Temporary medical centers (TMCs) provide healthcare services to victims of large-scale and catastrophic disasters. Examples of TMCs include  Red Cross medical tents, casualty collection points, and any temporary HCF established before the disaster to play a short term role in the immediate aftermath. TMC location has some stochastic characteristics similar to those we see in humanitarian logistics that should be considered. These include the possibility of TMCs being completely destroyed during the disaster, changes in the capacities and functionality of roads between TMCs and disaster sites, uncertainty in the fraction of usable medical supply in the aftermath,  and uncertainty in the size and location of demand.  In addition, helicopters are often used to transport medical supplies and people to the TMC and transport victims needing further care from TMCs to other HCFs. Thus, one should also consider helicopter-feasible locations. Despite the importance of TMC, as pointed out by \cite{ahmadi2017survey}, only a few papers addressed TMC location problems, and surprisingly under deterministic conditions. None of these studies incorporated equity objectives or constraints.



\section{Future Directions}\label{Sec:Future}

\noindent While great research efforts have been reported to improve stochastic HCF location theory, much work is still needed to incorporate equity and derive inequity-averse stochastic HCF location approaches and insights. We discuss a few critical research opportunities for the future in the following subsections.

\subsection{Analyzing Equity Measures}

\noindent The World Health Organization defines health equity as \textit{the absence of unfair and avoidable or remediable differences in health among population groups defined socially, economically, demographically, or geographically} (WHO 2021). Accordingly, when making HCF location--allocation decisions, one should ensure equal distributions of HCF, equal access to HCF, and equal utilization of healthcare services/HCFs across geographical areas, social groups, demographics, socioeconomic groups, and identified/non-identified groups under normal and abnormal (e.g., disaster, conflict, etc.) conditions.  However, to ensure this, one should first find the right measure of equity.  But, is the obvious objective the right one? Are the obvious and classical constraints the right ones? Is the impact of decisions obtained from classical models with the obvious objectives equitable when compared across various people or groups of people?

To date, we do not have a formal analysis of equity objectives and constraints under uncertainty. Thus, the first step toward developing tractable and realistic stochastic inequity-averse HCF location approaches is to rigorously analyze the mathematical similarities and differences between different mathematical representation of a measure of equity under uncertainty. This is to identify subsets of mathematically similar equity metrics and accordingly limit the scope of modeling and comparison (and later optimization) to the subset of mathematically different metrics. Such analysis will facilitate studying equity--equity trade-offs (based on distinct notions of equity arising from a choice of population grouping or equity metric) in addition to studying equity--efficiency trade-offs. Then, one can analyze the subset of different measures to identify their mathematical complexity (e.g., linearity, convexity, etc).  This will, in turn, determine the complexity of optimization models. For example, one might resort to a linear metric that yield solutions that are near-optimal rather than using a nonlinear and complex metric that yields optimal solutions (if the problem can even be solved).

\subsection{Capturing Uncertainty: Optimizer's Curse and Trade-offs}

\noindent  Acknowledging the inevitable uncertainty and the uncertainty--equity interaction in HCF location settings is crucial to devise inequity-averse models for emerging real-life HCF location problems. One possibility for hedging against uncertainty is to capture it in the parameters underlying optimization models to support the decision-making process.

 There are three main stochastic optimization (SO) frameworks: stochastic programming (SP), robust optimization (RO), and distributionally robust optimization (DRO). The main distinguishing feature between these frameworks concerns the knowledge of the probability distribution of the underlying random vector. Hence,  adopting one of these modeling frameworks depends mainly on the available information regarding uncertainty and its distribution.  If we know the uncertainty distribution or have sufficient and high-quality data to represent it, we resort to SP. In SP, we typically consider two- or multi-stage models in which the first-stage  (\textit{here-and-now}) decisions represent decisions to implement before any uncertainty is revealed. Then, in the subsequent stages, we often use a discrete number of scenarios (from the assumed known distribution) to capture the uncertainty and make \textit{recourse} (\textit{wait-and-see}) decisions that adapt to these scenarios.

The SP decision problem evaluates the objective and optimizes the decisions only for the given training sample (which may come from a biased distribution). Thus the resulting SP decisions may be overfitted or optimistically biased. In reality, possible future uncertainty realizations may differ from all training samples used in the optimization. Thus, SP solutions may demonstrate disappointing out-of-sample performance under the true distribution. Disappointing consequences in healthcare include, but are not limited to,  disparities in healthcare service and distribution of services, poor access to care, increased mortality and morbidity of the vulnerable population, and increased costs.  Mathematically, solutions to SP decision problems display an optimistic in-sample risk/outcome that cannot be realized out of sample. This phenomenon is known as the \textit{optimizer's curse} (i.e., optimization  based on imperfect estimates of distribution leads to biased decisions) and is reminiscent of overfitting effects in statistics \citep{esfahani2018data, smith2006optimizer}.  Unfortunately, as shown in Section~\ref{sec:EquityInHCF}, most existing studies employ SP or other ``sample-based'' approaches.

In contrast to SP, RO assumes that the decision-maker has no distributional knowledge about the underlying uncertainty, except for its support. RO seeks a particular measure of robustness against the uncertainty represented as \textit{deterministic variability} in the value of the problem's parameters. Specifically, in RO, we assume that uncertain parameters belong to an \textit{uncertainty set} of possible outcomes with some structure (e.g., ellipsoid or polyhedron). Then, optimization is based on the worst-case scenario occurring within this set, which inevitably leads to over-conservatism and suboptimal decisions for other more likely scenarios.

DRO is an alternative \textit{ambiguity-averse} approach to model uncertainty that offers a middle ground instead of the black-or-white view of the SP and RO approaches about knowing the distribution of uncertain parameters. In DRO, we model the distribution of uncertainty as a decision variable that belongs to an \textit{ambiguity set}, i.e., a family of all possible distribution of uncertainty that share some characteristics (e.g., mean, range, co-variance, etc). Optimization is then based on the distribution within this ambiguity set. 

DRO has recently become a promising approach for addressing real-life stochastic optimization problems due to the following striking benefits.  First, DRO models are more ``honest'' than their SP and RO counterparts as they acknowledge the presence of distributional uncertainty.  Various theoretical results show that DRO solutions faithfully mitigate the optimizer's curse and the associated disappointing out-of-sample consequences. Second, by allowing uncertainty to follow any distribution within the ambiguity set, DRO alleviates the unrealistic assumption of the decision-maker's complete knowledge of distributions.  Third, one can use intuitive and easy-to-approximate statistics (e.g., based on expert knowledge) to construct the ambiguity set. For example, we could estimate the average demand for primary care in each city of the Lehigh Valley in Pennsylvania (U.S.). Then, we can construct a mean--range ambiguity set of demand distributions, where the range is used to represent the error margin in the estimates that we seek protection against. Accordingly, a DRO model could optimize primary care clinic location decisions against all possible distributions in the ambiguity set that share these mean and range values. Some ambiguity sets also allow for incorporating the possibly small available data in the optimization while hedging against the estimation error of the associated empirical distribution (e.g., DRO using the Wasserstein metric \cite{esfahani2018data}).

Fourth, DRO models of some real-world problems are often more tractable  and have better performance than their SP counterparts (see, e.g., the following DRO facility location studies \cite{basciftci2019distributionally, luo2018distributionally, saif2020data,  ShehadehSanci,DMFRS, shehadeh2020distributionallyTucker, wang2020distributionally, wang2021two,wu2015approximation}). Fifth,  DRO may reflect decision-makers' ambiguity aversion and preference to err on the side of caution and seek robust location decisions to safeguard performance in adverse scenarios and mitigate direct and indirect costs of care and inequity (e.g., poor population health outcomes).  Finally, even when complete distributional information is available, DRO models often perform well. This suggests that it is worthwhile to consider DRO as an alternative approach for modeling uncertainty in HCF location.  \textit{Yet, despite the potential advantages, there is no inequity-averse DRO approach with equity and efficiency concerns for facility location within or outside healthcare applications.}

Indeed, there is a trade-off to using different approaches to model uncertainty that must also be considered when deriving inequity-averse HCF location models.  SP may provide an excellent basis if we unrealistically assume that we know the distributions of uncertain parameters. However, as we mentioned above, SP solutions often underdeliver and disappoint. In contrast, by focusing on hedging against ambiguity, DRO solutions may seem conservative and pessimistic. Thus, models that minimize the trade-off between considering distributional ambiguity (DRO pessimism) and following distributional belief (SP optimism) may offer a middle ground. \textit{ Unfortunately, this trade-off has not been studied within the context of facility location (except in \cite{shehadeh2020distributionallyTucker}) or in inequity-averse HCF location in particular. }

These gaps and challenges call for more efforts toward rigorous analyses of equity under different scenarios of information availability and ambiguity. Formal analyses are also needed to answer questions about \textit{when} to use each type of modeling approach (SP, DRO, or a trade-off between them) and about \textit{what} is the value of adopting each for inequity-averse HCF location. Extending classical models to derive distribution-free and tractable inequity-averse HCF location models is both theoretically and practically relevant. It will have worthwhile contributions to location science, optimization under uncertainity, and computational operations research.

\subsection{Dynamic Mapping and Databases}

\noindent  Location is the central element in all spatial analyses and varies in size and shape depending on the context. For many decades, locations or places played an important role in understanding health patterns, disease patterns, and healthcare service distributions. Historically, maps have been the primary source for storing and communicating spatial information. In the past two decades, the advancement in computer power and the emergence of  Geographic Information Systems (GIS; a system that captures, creates, stores, manages, analyzes, and maps all types of data) has allowed for better mapping and more widespread, complex, and comprehensive analyses than previously. GIS connects data to a map, integrating location data (so it is possible to know where things are) with all types of descriptive information (so it is possible to understand what things are like in a specific location).  Such advances have made it possible for governments, researchers, and others to seek answers to previously overly complex and unfeasible questions.

Data accuracy, correctness, and completeness are crucial elements affecting our ability to use GIS to analyze health and equity issues effectively. Unfortunately, despite the significant improvement in technologies to obtain geographic data, we often need to geocode specific data such as patient data (volume, disease, etc) and HCF utilization by different groups to undertake a particular geographical analysis.  As pointed out by \cite{lyseen2014review}, in the absence of standardized data collection methods or databases, this process imposes a significant challenge for health information systems in collecting data with adequate granularity, ensuring reliability and validity of the relevant health data, and maintaining appropriate privacy and security for the collected data. Moreover, \textit{even in a perfect situation when the needed data has been collected, it is often unavailable to researchers due to privacy laws regarding patients' medical information \cite{lyseen2014review}.}

Several developed countries and research groups create and publish interactive GIS maps of HCF locations with basic information. However, these are individual efforts; researchers and governments often work individually, and there are no integrated or standardized databases even within the same country.  Therefore, policies and strategies for developing standards-compliant and reliable databases on HCF and the ability to integrate these in GIS dynamically as information unfolds are important research directions and a prerequisite to developing data-driven inequity-averse HCF location models.  Such databases may include but are not limited to real-time data on the location, type, and characteristics of existing HCFs (location, scope of services, quality of service, types of insurance accepted, etc.),  utilization of existing HCFs (by, e.g., social and racial groups), characteristics (socioeconomic, education level, demographic, etc.) of people living in each geographical area (country, state, county, city, block, etc.), and demand and access for healthcare and health services across geographical areas and (social, economic, religious, racial, etc.). This information is essential for developing inequity-averse HCF location approaches that better mimic reality. 

A key challenge in collecting and using data for HCF location problems is the inevitability of bias in the data. A number of types of bias can arise, including {\em selection bias} (when  groups are over- or under-represented in the data compared to their true populations), {\em historical bias} (bias caused by events that occurred before the data collection or analysis), {\em confirmation bias} (researchers' tendencies to choose or trust data that supports their existing hypotheses), and {\em availability bias} (bias due to the use of data that are available, even if those are not the most relevant data). Bias in healthcare data have already been shown to produce negative outcomes (for example, an algorithm that interprets imaging results and that is trained on gender-biased data will produce worse outcomes for under-represented genders \cite{Larrazabal12592}), and it stands to reason that biased data used for HCF location studies will similarly lead to harms for some groups. See, e.g., \cite{himss,catalogofbias} for more on data bias.

\subsection{Multicriteria Approaches}

\noindent The specific approach or criteria to capture inequity are context-dependent and depend on the decision maker's perception of equity.  One can, of course, consider a multicriteria objective and a method to optimize multiple equity criteria (equity of access, socioeconomic equity, geographical equity) and efficiency criteria (e.g., transportation cost,  operational costs, etc). However, to the best of our knowledge, and according to a recent survey on HCF location \cite{ahmadi2017survey} and our analysis in Section~\ref{sec:EquityInHCF}, within the limited literature considering equity in stochastic HCF location, most studies only consider one equity-related policy objective, and only a few multi-objective models exist for the joint attainment of different equity objectives under uncertainty.

When we seek both equity  (e.g., equitable allocation of vaccination clinics across urban and rural areas) and efficiency (e.g., minimum operational cost),
there is the additional challenge of mathematically integrating them in a tractable model. As we mentioned earlier, considering equity may degrade efficiency. Thus, two fundamental questions arise in integrating equity and efficiency: \textit{how to regulate the trade-off between the two under uncertainty and potential distributional ambiguity}? and \textit{what is the price of equity (i.e., the efficiency difference between selecting an inequity-averse approach and not using an inequity-averse approach)?}


In theory,  multi-criteria optimization problems such as HCF location with equity and efficiency concerns can be modeled and solved using multi-objective optimization (MOO) approaches such as interactive approach, hierarchical approach, or a weighted objective function (see, e,g., \cite{ehrgott2005multicriteria} for a detailed discussion of MOO). However, multicriteria location models assume that decision-makers can articulate their objectives or have a well-defined metric for each objective. In most real-world applications such as HCF location,  the decision maker's priorities, goals, and equity concerns are not easy to articulate and could vary over time and among different decision-makers.    Moreover, we often need to incorporate the (potentially conflicting) perspective of many stakeholders. For example, locating hospitals is
a process that must take into consideration many different stakeholders \cite{burkey2012location}, including  \textit{patients who need access to the hospital, clinical staff who want an attractive and easy-to-reach workplace, taxpayers who want value for their dollars, politicians who want to demonstrate
their ability to deliver a better quality of life  and healthcare services, and more.}

Moreover, equity measures generally compare the impacts of actions (e.g., location decisions) on different groups and possibly weigh such effects based on the characteristics of a group (e.g., needs for a particular health service). Unfortunately, decision-makers cannot always quantify or predict HCF location decisions' direct/indirect impact on patient outcomes and access to care, among others. Thus, the cost associated with HCF location decisions is often subject to ambiguity and cannot be explicitly measured. Even in a perfect situation when we have a precise definition of equity and efficiency metrics, these metrics are often conflicting, and it is often challenging to arbitrate the conflict between them.  In addition, the incorporation of uncertainty increases the complexity of multi-criteria HCF location models.

\textit{These challenges call for more efforts toward bridging the gap between MOO (and other operations research) approaches, decision-makers' perspectives of equity and efficiency, and the actual cost or criteria used in practice.} The ability to quantify the trade-off between efficiency and equity would help decision-makers make informed and better location decisions \citep{karsu2015inequity}. In particular, to make decisions, a decision-maker needs to understand (a) what the efficiency loss might be and (b) what the equity loss might be for a specific solution. Although various studies analyze the price of fairness or price of equity (see, e.g., \cite{bertsimas2012efficiency, bertsimas2013fairness}) this price has not been adequately analyzed in HCF location contexts under uncertainty.

Analyzing the robustness of solutions obtained from different inequity-averse HCF location approaches under uncertainty is another venue for future research. Incorporating multiple equity measures and an analysis of the impact of uncertainty on each and the trade-off among them could shed some light on the question of how similar or different equity measures are. There are some attempts to analyze the commonality and differences of different equity measures in general facility location models \citep{batta2014public, karsu2015inequity, mulligan1991equality, lopez2003sum}. However, such analyses have not been conducted in the HCF location literature.

\subsection{Mobile HCF}

\noindent A mobile healthcare facility (MHCF) is a \textit{facility-like-vehicle} that can serve patients in a way similar to a static HCF when stationary but can also move from one place to another to provide health services to communities \citep{attipoe2020mobile, malone2020mobile, mcgowan2020mobile, DMFRS, santa2020mobile}. They offer a wide variety of  (prevention, testing, diagnostic) health services and are often staffed by a combination of physicians, nurses, community health workers, and other health professionals.   As longstanding community-based service delivery models, mobile facilities have the potential to help underserved communities overcome common barriers to accessing healthcare, including availability, time, geography, and trust \citep{clark2011lack, guruge2010immigrant, sommers2015health}, and have demonstrated improvements in health outcomes and reductions in cost \citep{brown2014mobile, oriol2009calculating, song2013mobile}.

MHCF also  offer an alternative healthcare delivery option when a disaster, conflict, or other event causes stationary HCFs (e.g., hospitals) to close or stop operations \citep{blackwell2007use, du2007mobile, fox1996costs,  gibson2011households}. For example, as of the writing of this book chapter, our world is still going through unprecedented times, fighting against the novel coronavirus (COVID-19) pandemic. Mobile clinics have been used to provide COVID-19 testing, vaccination, and triage services. Unfortunately, the deployment of such clinics has been associated with poor access and inequity.  According to data from GoodRx, in August 2020, 20\% of the U.S.\ population lived far from a COVID-19 testing clinic. Moreover, while some lived within 3 miles or less from a testing clinic, people in almost 17,000 census tracts had to drive an average of 22 miles to the nearest testing clinic. \textit{Classical stochastic mobile facility deployment, routing, and scheduling models  (see, e.g., \cite{halper2011mobile, DMFRS, lei2014multicut, lei2016two}) do not incorporate equity objectives or constraints and thus may produce such inequitable decisions}.

Developing inequity-averse MHCF location models is more challenging than the static HCF for the following reasons. First, there is limited literature on mobile facilities as compared to stationary facilities \citep{ahmadi2017survey}. Second,  in static HCF problems, we usually consider opening facilities at fixed locations. In contrast, MHCF location problems have the added challenge of determining a \textit{routing plan} and a \textit{time schedule} for each MHCF in the fleet (i.e., the node that each MHCF is located at in each time period). Mathematically, stochastic routing and scheduling are challenging stochastic combinatorial optimization problems. Thus, \textit{integrated MHCF deployment (e.g., determining the number of mobile vaccination clinics) and their capacity, routing, and scheduling problems with equity and efficiency criteria under uncertainty represent a new class of complex combinatorial, multicriteria, stochastic optimization problems}. To date, optimization (or other) tools have not been employed to analyze and address these problems.  Thus, the development of tractable inequity-averse mobile facility deployment, location, routing, and scheduling is practically important to support decisions in all applications of mobile HCF and technically interesting.

\subsection{Robust and Resilient HCF locations}

\noindent The supply and demand for healthcare are often not balanced. In addition, multiple disruptive events can shock the healthcare network and exaggerate the imbalance between supply and demand. Long-term disruptions including disasters,  infectious disease outbreaks, and economic and political conflicts can cause significant shocks to HCF operations over a long period. Unfortunately,  vulnerable populations often suffer the most during such a disruption.  Thus, we need strategies for building a robust and resilient inequity-averse HCF location framework that can adjust to changes caused by long-term disruptions. Resilience considers the healthcare network's vulnerability to disruption, i.e., the system's ability to function even if one or more nodes (HCFs) are removed from the network or disabled (e.g., a temporary or permanent closure of a hospital or a halt in operations).

Furthermore, developing models for inequity-averse HCF location incorporating the dynamic unfolding of disruptive events is practically relevant and theoretically interesting. The emerging complex healthcare challenges also demand new models that account for equity when pursuing actions such as reorganizing existing HCF networks, which may involve relocation or closure of existing facilities,  redistribution of capacity,
and mergers of facilities.


\section{Conclusion}

\noindent Healthcare and HCF location problems have become noticeably more essential to society in the past decade due to the global decrease in birth rates, higher longevity and associated growth in the elderly population, outspread of chronic and infectious diseases, increase in environmental and economic problems, conflicts, and disasters, and the associated impact on health and healthcare supply and demand.  These globally pervasive trends continue to create new challenges and equity concerns that classical HCF location models cannot address.

In this chapter, we focused on the issue of equity in stochastic HCF locations. We recognize that uncertainty is an intrinsic property of HCF location and explore how uncertainty and equity interact in ways that should not be ignored. The primary goal is to bring attention to equity issues and provide an understanding of the existing recent effort to incorporate equity measures in stochastic HCF locations. Our key findings are

\begin{itemize}
\item There is limited literature considering equity in HCF location and equity--uncertainty interaction.

\item Within the limited literature, most studies use the distance that a patient must travel or the travel time to HCF to measure equity and accessibility, with minimizing the maximum distance to the nearest facility as the most frequent measure. Some other studies account for factors such as spatial accessibility, multi-modal travel, temporal service availability, competition, multiple and hierarchical services,  socio-economic and personal factors, etc.\ in order to more realistically reproduce users' behaviors \citep{bruno2020improving, higgs2019using,jin2019evaluating,mathon2018cross, mayaud2019future, lin2018multi,shin2018improving, yin2018inequality}. 

\item Most studies, especially in EMS, use covering-based models to ensure equity. However, covering-based models favor locating HCFs in more densely populated areas, potentially resulting in longer travel or response times for patients in more rural areas. In addition, traditional covering models may lead to solutions in which the coverage pattern is quite good for those nodes counted as covered but extremely poor for those not covered.

\item Some studies investigate the trade-off between one or more equity objectives and multiple efficiency objectives.  However, no study has analyzed the equity--equity trade-off (i.e., the trade-off between equity metrics) or the equity--efficiency trade-off with multiple equity metrics.

\item  Stochastic optimization (SO) models with \textit{recourse} have been adopted widely to hedge against uncertainty, particularly in two-stage settings.  Recourse models result when (first-stage) decisions must be fixed before information relevant to uncertainty is available, and some (second-stage) recourse decisions are delayed until this information is available \citep{birge2011introduction, higle2005stochastic}. Thus, recourse decisions offer an opportunity to adjust to the received information about the uncertain data.  The two-stage structure of stochastic HCF location problems (e.g., choose number and locations of facilities before we know the future demand and take actions once demand uncertainty has been revealed, for example, by assigning customers to facilities) has made them attractive to be modeled using recourse models \citep{ahmadi2017survey,snyder2006facility}.

\item  Existing SO models for HCF location assumes that the distribution of uncertainty is fully known or there is sufficient data to model it. To date, there are no data-driven and distribution-free inequity-averse HCF location models.

\item While there is extensive literature on static facility location problems, there is limited literature on stochastic mobile facility locations. To date,  there are no inequity-averse mobile healthcare facility deployment, routing, and scheduling models.

\item No studies rigorously analyze the mathematical similarities, differences, and complexity of equity objectives and constraints. In addition, there are no standardized and open-access databases for HCF locations that incorporate all the geographical, socioeconomic, and other relevant data that one needs to derive and test inequity-averse models. 

\end{itemize}

We discuss and provide new directions for future research opportunities by recognizing these challenges and gaps in the literature. Our main recommendations for future research include:
    \begin{itemize}
        \item Conducting formal analyses of equity metrics and their mathematical complexity under uncertainty and distributional ambiguity
        \item Developing data-driven, distribution-free, and tractable multi-criteria  inequity-averse static and mobile HCF location approaches
        \item Developing standardized, granular, and dynamic HCF databases and integrate these with recent GIS technology
        \item Developing robust and resilient inequity-averse stochastic HCF location approaches
    \end{itemize}

\vspace{2mm}

\noindent \textbf{References}

\bibliographystyle{elsarticle-harv}
\bibliography{Equity_SHCFL}

\begin{thebibliography}{117}
\expandafter\ifx\csname natexlab\endcsname\relax\def\natexlab#1{#1}\fi
\expandafter\ifx\csname url\endcsname\relax
  \def\url#1{\texttt{#1}}\fi
\expandafter\ifx\csname urlprefix\endcsname\relax\def\urlprefix{URL }\fi

\bibitem[{Aboolian et~al.(2016)Aboolian, Berman, and
  Verter}]{aboolian2016maximal}
Aboolian, R., Berman, O., Verter, V., 2016. Maximal accessibility network
  design in the public sector. Transportation Science 50~(1), 336--347.

\bibitem[{Ahmadi-Javid et~al.(2017)Ahmadi-Javid, Seyedi, and
  Syam}]{ahmadi2017survey}
Ahmadi-Javid, A., Seyedi, P., Syam, S.~S., 2017. A survey of healthcare
  facility location. Computers \& Operations Research 79, 223--263.

\bibitem[{Ares et~al.(2016)Ares, De~Vries, and Huisman}]{ares2016column}
Ares, J.~N., De~Vries, H., Huisman, D., 2016. A column generation approach for
  locating roadside clinics in africa based on effectiveness and equity.
  European Journal of Operational Research 254~(3), 1002--1016.

\bibitem[{Attipoe-Dorcoo et~al.(2020)Attipoe-Dorcoo, Delgado, Gupta, Bennet,
  Oriol, and Jain}]{attipoe2020mobile}
Attipoe-Dorcoo, S., Delgado, R., Gupta, A., Bennet, J., Oriol, N.~E., Jain,
  S.~H., 2020. Mobile health clinic model in the covid-19 pandemic: lessons
  learned and opportunities for policy changes and innovation. International
  Journal for Equity in Health 19~(1), 1--5.

\bibitem[{Basciftci et~al.(2020)Basciftci, Ahmed, and
  Shen}]{basciftci2019distributionally}
Basciftci, B., Ahmed, S., Shen, S., 2020. Distributionally robust facility
  location problem under decision-dependent stochastic demand. European Journal
  of Operational Research.

\bibitem[{Batta et~al.(2014)Batta, Lejeune, and Prasad}]{batta2014public}
Batta, R., Lejeune, M., Prasad, S., 2014. Public facility location using
  dispersion, population, and equity criteria. European Journal of Operational
  Research 234~(3), 819--829.

\bibitem[{Beheshtifar and Alimoahmmadi(2015)}]{beheshtifar2015multiobjective}
Beheshtifar, S., Alimoahmmadi, A., 2015. A multiobjective optimization approach
  for location-allocation of clinics. International Transactions in Operational
  Research 22~(2), 313--328.

\bibitem[{Beraldi and Bruni(2009)}]{beraldi2009probabilistic}
Beraldi, P., Bruni, M.~E., 2009. A probabilistic model applied to emergency
  service vehicle location. European Journal of Operational Research 196~(1),
  323--331.

\bibitem[{Beraldi et~al.(2004)Beraldi, Bruni, and
  Conforti}]{beraldi2004designing}
Beraldi, P., Bruni, M.~E., Conforti, D., 2004. Designing robust emergency
  medical service via stochastic programming. European Journal of Operational
  Research 158~(1), 183--193.

\bibitem[{Bertsimas et~al.(2012)Bertsimas, Farias, and
  Trichakis}]{bertsimas2012efficiency}
Bertsimas, D., Farias, V.~F., Trichakis, N., 2012. On the efficiency-fairness
  trade-off. Management Science 58~(12), 2234--2250.

\bibitem[{Bertsimas et~al.(2013)Bertsimas, Farias, and
  Trichakis}]{bertsimas2013fairness}
Bertsimas, D., Farias, V.~F., Trichakis, N., 2013. Fairness, efficiency, and
  flexibility in organ allocation for kidney transplantation. Operations
  Research 61~(1), 73--87.

\bibitem[{Birge and Louveaux(2011)}]{birge2011introduction}
Birge, J.~R., Louveaux, F., 2011. Introduction to stochastic programming.
  Springer Science \& Business Media.

\bibitem[{Blackwell and Bosse(2007)}]{blackwell2007use}
Blackwell, T., Bosse, M., 2007. Use of an innovative design mobile hospital in
  the medical response to hurricane katrina. Annals of emergency medicine
  49~(5), 580--588.

\bibitem[{Brown-Connolly et~al.(2014)Brown-Connolly, Concha, and
  English}]{brown2014mobile}
Brown-Connolly, N.~E., Concha, J.~B., English, J., 2014. Mobile health is worth
  it! economic benefit and impact on health of a population-based mobile
  screening program in new mexico. Telemedicine and e-Health 20~(1), 18--23.

\bibitem[{Bruno et~al.(2020)Bruno, Cavola, Diglio, and
  Piccolo}]{bruno2020improving}
Bruno, G., Cavola, M., Diglio, A., Piccolo, C., 2020. Improving spatial
  accessibility to regional health systems through facility capacity
  management. Socio-Economic Planning Sciences 71, 100881.

\bibitem[{Burkey et~al.(2012)Burkey, Bhadury, and Eiselt}]{burkey2012location}
Burkey, M.~L., Bhadury, J., Eiselt, H.~A., 2012. A location-based comparison of
  health care services in four us states with efficiency and equity.
  Socio-Economic Planning Sciences 46~(2), 157--163.

\bibitem[{Cardoso et~al.(2015)Cardoso, Oliveira, Barbosa-P{\'o}voa, and
  Nickel}]{cardoso2015integrated}
Cardoso, T., Oliveira, M.~D., Barbosa-P{\'o}voa, A., Nickel, S., 2015. An
  integrated approach for planning a long-term care network with uncertainty,
  strategic policy and equity considerations. European Journal of Operational
  Research 247~(1), 321--334.

\bibitem[{Cardoso et~al.(2016)Cardoso, Oliveira, Barbosa-P{\'o}voa, and
  Nickel}]{cardoso2016moving}
Cardoso, T., Oliveira, M.~D., Barbosa-P{\'o}voa, A., Nickel, S., 2016. Moving
  towards an equitable long-term care network: A multi-objective and
  multi-period planning approach. Omega 58, 69--85.

\bibitem[{{Catalog of Bias}(2021)}]{catalogofbias}
{Catalog of Bias}, 2021. Uncovering and removing data bias in healthcare.
\newline\urlprefix\url{https://catalogofbias.org/}

\bibitem[{Chakravarty(1999)}]{chakravarty1999measuring}
Chakravarty, S.~R., 1999. Measuring inequality: the axiomatic approach. In:
  Handbook of income inequality measurement. Springer, pp. 163--186.

\bibitem[{Chang et~al.(2006)Chang, Lee, and Kim}]{chang2006optimal}
Chang, K.-N., Lee, K.-D., Kim, D., 2006. Optimal timeslot and channel
  allocation considering fairness for multicell cdma/tdd systems. Computers \&
  operations research 33~(11), 3203--3218.

\bibitem[{Chanta et~al.(2011)Chanta, Mayorga, Kurz, and
  McLay}]{chanta2011minimum}
Chanta, S., Mayorga, M.~E., Kurz, M.~E., McLay, L.~A., 2011. The minimum p-envy
  location problem: a new model for equitable distribution of emergency
  resources. IIE Transactions on Healthcare Systems Engineering 1~(2),
  101--115.

\bibitem[{Chanta et~al.(2014)Chanta, Mayorga, and McLay}]{chanta2014improving}
Chanta, S., Mayorga, M.~E., McLay, L.~A., 2014. Improving emergency service in
  rural areas: a bi-objective covering location model for ems systems. Annals
  of Operations Research 221~(1), 133--159.

\bibitem[{Ciss{\'e} et~al.(2017)Ciss{\'e}, Yal{\c{c}}{\i}nda{\u{g}}, Kergosien,
  {\c{S}}ahin, Lent{\'e}, and Matta}]{cisse2017or}
Ciss{\'e}, M., Yal{\c{c}}{\i}nda{\u{g}}, S., Kergosien, Y., {\c{S}}ahin, E.,
  Lent{\'e}, C., Matta, A., 2017. Or problems related to home health care: A
  review of relevant routing and scheduling problems. Operations Research for
  Health Care 13, 1--22.

\bibitem[{Clark et~al.(2011)Clark, Soukup, Govindarajulu, Riden, Tovar, and
  Johnson}]{clark2011lack}
Clark, C.~R., Soukup, J., Govindarajulu, U., Riden, H.~E., Tovar, D.~A.,
  Johnson, P.~A., 2011. Lack of access due to costs remains a problem for some
  in massachusetts despite the state’s health reforms. Health Affairs 30~(2),
  247--255.

\bibitem[{Daskin(1982)}]{Daskin1982}
Daskin, M.~S., 1982. Application of an expected covering model to emergency
  medical service system design. Decision Sciences 13, 416--439.

\bibitem[{Daskin(1983)}]{Daskin1983}
Daskin, M.~S., 1983. A maximum expected covering location model: Formulation,
  properties and heuristic solution. Transportation Science 17~(1), 48--70.

\bibitem[{Daskin and Dean(2005)}]{daskin2005location}
Daskin, M.~S., Dean, L.~K., 2005. Location of health care facilities.
  Operations research and health care, 43--76.

\bibitem[{Du~Mortier and Coninx(2007)}]{du2007mobile}
Du~Mortier, S., Coninx, R., 2007. Mobile health units in emergency operations:
  a methodological approach. Humanitarian Practice Network, Overseas
  Development Inst.

\bibitem[{Ehrgott(2005)}]{ehrgott2005multicriteria}
Ehrgott, M., 2005. Multicriteria optimization. Vol. 491. Springer Science \&
  Business Media.

\bibitem[{Enayati et~al.(2019)Enayati, Mayorga, Toro-D{\'\i}az, and
  Albert}]{enayati2019identifying}
Enayati, S., Mayorga, M.~E., Toro-D{\'\i}az, H., Albert, L.~A., 2019.
  Identifying trade-offs in equity and efficiency for simultaneously optimizing
  location and multipriority dispatch of ambulances. International Transactions
  in Operational Research 26~(2), 415--438.

\bibitem[{Esfahani and Kuhn(2018)}]{esfahani2018data}
Esfahani, P.~M., Kuhn, D., 2018. Data-driven distributionally robust
  optimization using the wasserstein metric: Performance guarantees and
  tractable reformulations. Mathematical Programming 171~(1-2), 115--166.

\bibitem[{Fahimnia et~al.(2017)Fahimnia, Jabbarzadeh, Ghavamifar, and
  Bell}]{fahimnia2017supply}
Fahimnia, B., Jabbarzadeh, A., Ghavamifar, A., Bell, M., 2017. Supply chain
  design for efficient and effective blood supply in disasters. International
  Journal of Production Economics 183, 700--709.

\bibitem[{Fox-Rushby and Foord(1996)}]{fox1996costs}
Fox-Rushby, J.~A., Foord, F., 1996. Costs, effects and cost-effectiveness
  analysis of a mobile maternal health care service in west kiang, the gambia.
  Health Policy 35~(2), 123--143.

\bibitem[{Gendreau et~al.(2006)Gendreau, Laporte, and
  Semet}]{gendreau2006maximal}
Gendreau, M., Laporte, G., Semet, F., 2006. The maximal expected coverage
  relocation problem for emergency vehicles. Journal of the Operational
  Research Society 57~(1), 22--28.

\bibitem[{Gibson et~al.(2011)Gibson, Deng, Boe-Gibson, Rozelle, and
  Huang}]{gibson2011households}
Gibson, J., Deng, X., Boe-Gibson, G., Rozelle, S., Huang, J., 2011. Which
  households are most distant from health centers in rural china? evidence from
  a gis network analysis. GeoJournal 76~(3), 245--255.

\bibitem[{Grieco et~al.(2020)Grieco, Utley, and Crowe}]{grieco2020operational}
Grieco, L., Utley, M., Crowe, S., 2020. Operational research applied to
  decisions in home health care: A systematic literature review. Journal of the
  Operational Research Society, 1--32.

\bibitem[{G{\"u}ne{\c{s}} et~al.(2019)G{\"u}ne{\c{s}}, Melo, and
  Nickel}]{gunecs2019location}
G{\"u}ne{\c{s}}, E.~D., Melo, T., Nickel, S., 2019. Location problems in
  healthcare. In: Location science. Springer, pp. 657--686.

\bibitem[{G{\"u}ne{\c{s}} et~al.(2014)G{\"u}ne{\c{s}}, Yaman, {\c{C}}ekyay, and
  Verter}]{gunecs2014matching}
G{\"u}ne{\c{s}}, E.~D., Yaman, H., {\c{C}}ekyay, B., Verter, V., 2014. Matching
  patient and physician preferences in designing a primary care facility
  network. Journal of the Operational Research Society 65~(4), 483--496.

\bibitem[{Guruge et~al.(2010)Guruge, Hunter, Barker, McNally, and
  Magalhaes}]{guruge2010immigrant}
Guruge, S., Hunter, J., Barker, K., McNally, M.~J., Magalhaes, L., 2010.
  Immigrant women’s experiences of receiving care in a mobile health clinic.
  Journal of Advanced Nursing 66~(2), 350--359.

\bibitem[{Guti{\'e}rrez and Vidal(2013)}]{gutierrez2013home}
Guti{\'e}rrez, E.~V., Vidal, C.~J., 2013. Home health care logistics management
  problems: A critical review of models and methods. Revista Facultad de
  Ingenier{\'\i}a Universidad de Antioquia~(68), 160--175.

\bibitem[{Gutjahr and Fischer(2018)}]{gutjahr2018equity}
Gutjahr, W.~J., Fischer, S., 2018. Equity and deprivation costs in humanitarian
  logistics. European Journal of Operational Research 270~(1), 185--197.

\bibitem[{Halper and Raghavan(2011)}]{halper2011mobile}
Halper, R., Raghavan, S., 2011. The mobile facility routing problem.
  Transportation Science 45~(3), 413--434.

\bibitem[{Hawthorne and Kwan(2013)}]{hawthorne2013exploring}
Hawthorne, T.~L., Kwan, M.-P., 2013. Exploring the unequal landscapes of
  healthcare accessibility in lower-income urban neighborhoods through
  qualitative inquiry. Geoforum 50, 97--106.

\bibitem[{{Healthcare Information and Management Systems Society
  (HIMSS)}(2021)}]{himss}
{Healthcare Information and Management Systems Society (HIMSS)}, 2021.
  Uncovering and removing data bias in healthcare.
\newline\urlprefix\url{https://www.himss.org/resources/uncovering-and-removing-data-bias-healthcare}

\bibitem[{Higgs et~al.(2019)Higgs, Langford, Jarvis, Page, Richards, and
  Fry}]{higgs2019using}
Higgs, G., Langford, M., Jarvis, P., Page, N., Richards, J., Fry, R., 2019.
  Using geographic information systems to investigate variations in
  accessibility to ‘extended hours’ primary healthcare provision. Health \&
  social care in the community 27~(4), 1074--1084.

\bibitem[{Higle(2005)}]{higle2005stochastic}
Higle, J.~L., 2005. Stochastic programming: Optimization when uncertainty
  matters. In: Emerging Theory, Methods, and Applications. Informs, pp. 30--53.

\bibitem[{Ingolfsson et~al.(2008)Ingolfsson, Budge, and
  Erkut}]{ingolfsson2008optimal}
Ingolfsson, A., Budge, S., Erkut, E., 2008. Optimal ambulance location with
  random delays and travel times. Health Care management science 11~(3),
  262--274.

\bibitem[{Jabbarzadeh et~al.(2014)Jabbarzadeh, Fahimnia, and
  Seuring}]{jabbarzadeh2014dynamic}
Jabbarzadeh, A., Fahimnia, B., Seuring, S., 2014. Dynamic supply chain network
  design for the supply of blood in disasters: A robust model with real world
  application. Transportation Research Part E: Logistics and Transportation
  Review 70, 225--244.

\bibitem[{Jin et~al.(2015)Jin, Cheng, Lu, Huang, and Cao}]{jin2015spatial}
Jin, C., Cheng, J., Lu, Y., Huang, Z., Cao, F., 2015. Spatial inequity in
  access to healthcare facilities at a county level in a developing country: a
  case study of deqing county, zhejiang, china. International journal for
  equity in health 14~(1), 1--21.

\bibitem[{Jin et~al.(2019)Jin, Liu, Tong, Gong, and Liu}]{jin2019evaluating}
Jin, M., Liu, L., Tong, D., Gong, Y., Liu, Y., 2019. Evaluating the spatial
  accessibility and distribution balance of multi-level medical service
  facilities. International journal of environmental research and public health
  16~(7), 1150.

\bibitem[{Karsu and Morton(2015)}]{karsu2015inequity}
Karsu, {\"O}., Morton, A., 2015. Inequity averse optimization in operational
  research. European journal of operational research 245~(2), 343--359.

\bibitem[{Khodaparasti et~al.(2016)Khodaparasti, Maleki, Bruni, Jahedi,
  Beraldi, and Conforti}]{khodaparasti2016balancing}
Khodaparasti, S., Maleki, H.~R., Bruni, M.~E., Jahedi, S., Beraldi, P.,
  Conforti, D., 2016. Balancing efficiency and equity in location-allocation
  models with an application to strategic ems design. Optimization Letters
  10~(5), 1053--1070.

\bibitem[{Kim et~al.(2015)Kim, Pasupathy, and Henderson}]{kim2015guide}
Kim, S., Pasupathy, R., Henderson, S.~G., 2015. A guide to sample average
  approximation. In: Handbook of simulation optimization. Springer, pp.
  207--243.

\bibitem[{Kleywegt et~al.(2002)Kleywegt, Shapiro, and Homem-de
  Mello}]{kleywegt2002sample}
Kleywegt, A.~J., Shapiro, A., Homem-de Mello, T., 2002. The sample average
  approximation method for stochastic discrete optimization. SIAM Journal on
  Optimization 12~(2), 479--502.

\bibitem[{Kostreva et~al.(2004)Kostreva, Ogryczak, and
  Wierzbicki}]{kostreva2004equitable}
Kostreva, M.~M., Ogryczak, W., Wierzbicki, A., 2004. Equitable aggregations and
  multiple criteria analysis. European Journal of Operational Research 158~(2),
  362--377.

\bibitem[{Larrazabal et~al.(2020)Larrazabal, Nieto, Peterson, Milone, and
  Ferrante}]{Larrazabal12592}
Larrazabal, A.~J., Nieto, N., Peterson, V., Milone, D.~H., Ferrante, E., 2020.
  Gender imbalance in medical imaging datasets produces biased classifiers for
  computer-aided diagnosis. Proceedings of the National Academy of Sciences
  117~(23), 12592--12594.

\bibitem[{Lei et~al.(2014)Lei, Lin, and Miao}]{lei2014multicut}
Lei, C., Lin, W.-H., Miao, L., 2014. A multicut l-shaped based algorithm to
  solve a stochastic programming model for the mobile facility routing and
  scheduling problem. European Journal of operational research 238~(3),
  699--710.

\bibitem[{Lei et~al.(2016)Lei, Lin, and Miao}]{lei2016two}
Lei, C., Lin, W.-H., Miao, L., 2016. A two-stage robust optimization approach
  for the mobile facility fleet sizing and routing problem under uncertainty.
  Computers \& Operations Research 67, 75--89.

\bibitem[{Levinson(2010)}]{levinson2010equity}
Levinson, D., 2010. Equity effects of road pricing: A review. Transport Reviews
  30~(1), 33--57.

\bibitem[{Li et~al.(2011)Li, Zhao, Zhu, and Wyatt}]{li2011covering}
Li, X., Zhao, Z., Zhu, X., Wyatt, T., 2011. Covering models and optimization
  techniques for emergency response facility location and planning: a review.
  Mathematical Methods of Operations Research 74~(3), 281--310.

\bibitem[{Lin et~al.(2018)Lin, Wan, Sheets, Gong, and Davies}]{lin2018multi}
Lin, Y., Wan, N., Sheets, S., Gong, X., Davies, A., 2018. A multi-modal
  relative spatial access assessment approach to measure spatial accessibility
  to primary care providers. International journal of health geographics
  17~(1), 1--22.

\bibitem[{L{\'o}pez-De-Los-Mozos and Mesa(2003)}]{lopez2003sum}
L{\'o}pez-De-Los-Mozos, M.~C., Mesa, J.~A., 2003. The sum of absolute
  differences on a network: algorithm and comparison with other equality
  measures. INFOR: Information Systems and Operational Research 41~(2),
  195--210.

\bibitem[{Luo and Mehrotra(2018)}]{luo2018distributionally}
Luo, F., Mehrotra, S., 2018. Distributionally robust optimization with decision
  dependent ambiguity sets. arXiv preprint arXiv:1806.09215.

\bibitem[{Lyseen et~al.(2014)Lyseen, N{\o}hr, S{\o}rensen, Gudes, Geraghty,
  Shaw, Bivona-Tellez, Group, et~al.}]{lyseen2014review}
Lyseen, A.-K., N{\o}hr, C., S{\o}rensen, E.-M., Gudes, O., Geraghty, E., Shaw,
  N.~T., Bivona-Tellez, C., Group, I. H. G.~W., et~al., 2014. A review and
  framework for categorizing current research and development in health related
  geographical information systems (gis) studies. Yearbook of medical
  informatics 23~(01), 110--124.

\bibitem[{Mak et~al.(1999)Mak, Morton, and Wood}]{mak1999monte}
Mak, W.-K., Morton, D.~P., Wood, R.~K., 1999. {M}onte {C}arlo bounding
  techniques for determining solution quality in stochastic programs.
  Operations research letters 24~(1), 47--56.

\bibitem[{Malone et~al.(2020)Malone, Williams, Fawzi, Bennet, Hill, Katz, and
  Oriol}]{malone2020mobile}
Malone, N.~C., Williams, M.~M., Fawzi, M. C.~S., Bennet, J., Hill, C., Katz,
  J.~N., Oriol, N.~E., 2020. Mobile health clinics in the united states.
  International journal for equity in health 19~(1), 1--9.

\bibitem[{Mar{\'\i}n et~al.(2010)Mar{\'\i}n, Nickel, and
  Velten}]{marin2010extended}
Mar{\'\i}n, A., Nickel, S., Velten, S., 2010. An extended covering model for
  flexible discrete and equity location problems. Mathematical Methods of
  Operations Research 71~(1), 125--163.

\bibitem[{Mathon et~al.(2018)Mathon, Apparicio, and
  Lachapelle}]{mathon2018cross}
Mathon, D., Apparicio, P., Lachapelle, U., 2018. Cross-border spatial
  accessibility of health care in the north-east department of haiti.
  International journal of health geographics 17~(1), 1--15.

\bibitem[{Mayaud et~al.(2019)Mayaud, Tran, Pereira, and
  Nuttall}]{mayaud2019future}
Mayaud, J.~R., Tran, M., Pereira, R.~H., Nuttall, R., 2019. Future access to
  essential services in a growing smart city: The case of surrey, british
  columbia. Computers, Environment and Urban Systems 73, 1--15.

\bibitem[{McCoy and Lee(2014)}]{mccoy2014using}
McCoy, J.~H., Lee, H.~L., 2014. Using fairness models to improve equity in
  health delivery fleet management. Production and Operations Management
  23~(6), 965--977.

\bibitem[{McGowan et~al.(2020)McGowan, Baxter, Deola, Gayford, Marston,
  Cummings, and Checchi}]{mcgowan2020mobile}
McGowan, C.~R., Baxter, L., Deola, C., Gayford, M., Marston, C., Cummings, R.,
  Checchi, F., 2020. Mobile clinics in humanitarian emergencies: a systematic
  review. Conflict and health 14~(1), 4.

\bibitem[{McLay(2009)}]{mclay2009maximum}
McLay, L.~A., 2009. A maximum expected covering location model with two types
  of servers. IIE Transactions 41~(8), 730--741.

\bibitem[{Mestre et~al.(2015)Mestre, Oliveira, and
  Barbosa-P{\'o}voa}]{mestre2015location}
Mestre, A.~M., Oliveira, M.~D., Barbosa-P{\'o}voa, A.~P., 2015.
  Location--allocation approaches for hospital network planning under
  uncertainty. European Journal of Operational Research 240~(3), 791--806.

\bibitem[{Mitropoulos et~al.(2006)Mitropoulos, Mitropoulos, Giannikos, and
  Sissouras}]{mitropoulos2006biobjective}
Mitropoulos, P., Mitropoulos, I., Giannikos, I., Sissouras, A., 2006. A
  biobjective model for the locational planning of hospitals and health
  centers. Health Care Management Science 9~(2), 171--179.

\bibitem[{Mostajabdaveh et~al.(2019)Mostajabdaveh, Gutjahr, and
  Sibel~Salman}]{mostajabdaveh2019inequity}
Mostajabdaveh, M., Gutjahr, W.~J., Sibel~Salman, F., 2019. Inequity-averse
  shelter location for disaster preparedness. IISE Transactions 51~(8),
  809--829.

\bibitem[{Mulligan(1991)}]{mulligan1991equality}
Mulligan, G.~F., 1991. Equality measures and facility location. Papers in
  Regional Science 70~(4), 345--365.

\bibitem[{Naoum-Sawaya and Elhedhli(2013)}]{naoum2013stochastic}
Naoum-Sawaya, J., Elhedhli, S., 2013. A stochastic optimization model for
  real-time ambulance redeployment. Computers \& Operations Research 40~(8),
  1972--1978.

\bibitem[{Noyan(2010)}]{noyan2010alternate}
Noyan, N., 2010. Alternate risk measures for emergency medical service system
  design. Annals of Operations Research 181~(1), 559--589.

\bibitem[{Oliveira and Bevan(2006)}]{oliveira2006modelling}
Oliveira, M.~D., Bevan, G., 2006. Modelling the redistribution of hospital
  supply to achieve equity taking account of patient's behaviour. Health care
  management science 9~(1), 19--30.

\bibitem[{Oriol et~al.(2009)Oriol, Cote, Vavasis, Bennet, DeLorenzo, Blanc, and
  Kohane}]{oriol2009calculating}
Oriol, N.~E., Cote, P.~J., Vavasis, A.~P., Bennet, J., DeLorenzo, D., Blanc,
  P., Kohane, I., 2009. Calculating the return on investment of mobile
  healthcare. BMC medicine 7~(1), 1--6.

\bibitem[{Panzera and Postiglione(2020)}]{panzera2020measuring}
Panzera, D., Postiglione, P., 2020. Measuring the spatial dimension of regional
  inequality: An approach based on the gini correlation measure. Social
  Indicators Research 148~(2), 379--394.

\bibitem[{Rahman and Smith(2000)}]{rahman2000use}
Rahman, S.-u., Smith, D.~K., 2000. Use of location-allocation models in health
  service development planning in developing nations. European Journal of
  Operational Research 123~(3), 437--452.

\bibitem[{Rahmaniani et~al.(2014)Rahmaniani, Rahmaniani, and
  Jabbarzadeh}]{rahmaniani2014variable}
Rahmaniani, R., Rahmaniani, G., Jabbarzadeh, A., 2014. Variable neighborhood
  search based evolutionary algorithm and several approximations for balanced
  location--allocation design problem. The International Journal of Advanced
  Manufacturing Technology 72~(1-4), 145--159.

\bibitem[{Rajagopalan and Saydam(2009)}]{rajagopalan2009minimum}
Rajagopalan, H.~K., Saydam, C., 2009. A minimum expected response model:
  Formulation, heuristic solution, and application. Socio-Economic Planning
  Sciences 43~(4), 253--262.

\bibitem[{Rajagopalan et~al.(2011)Rajagopalan, Saydam, Setzler, Sharer,
  et~al.}]{rajagopalan2011ambulance}
Rajagopalan, H.~K., Saydam, C., Setzler, H., Sharer, E., et~al., 2011.
  Ambulance deployment and shift scheduling: An integrated approach. Journal of
  Service Science and Management 4~(01), 66.

\bibitem[{Rajagopalan et~al.(2008)Rajagopalan, Saydam, and
  Xiao}]{rajagopalan2008multiperiod}
Rajagopalan, H.~K., Saydam, C., Xiao, J., 2008. A multiperiod set covering
  location model for dynamic redeployment of ambulances. Computers \&
  Operations Research 35~(3), 814--826.

\bibitem[{Rawls(1999)}]{rawls1999theory}
Rawls, J., 1999. A theory of justice: Revised edition. Harvard university
  press.

\bibitem[{Saif and Delage(2020)}]{saif2020data}
Saif, A., Delage, E., 2020. Data-driven distributionally robust capacitated
  facility location problem. European Journal of Operational Research.

\bibitem[{Santa~Gonz{\'a}lez et~al.(2020)Santa~Gonz{\'a}lez, Cherkesly,
  Crainic, and Rancourt}]{santa2020mobile}
Santa~Gonz{\'a}lez, R., Cherkesly, M., Crainic, T.~G., Rancourt, M.-{\`E}.,
  2020. Mobile clinics deployment for humanitarian relief: A multi-period
  location-routing problem.

\bibitem[{Saveh-Shemshaki et~al.(2012)Saveh-Shemshaki, Shechter, Tang, and
  Isaac-Renton}]{saveh2012setting}
Saveh-Shemshaki, F., Shechter, S., Tang, P., Isaac-Renton, J., 2012. Setting
  sites for faster results: Optimizing locations and capacities of new
  tuberculosis testing laboratories. IIE Transactions on Healthcare Systems
  Engineering 2~(4), 248--258.

\bibitem[{Shapiro et~al.(2021)Shapiro, Dentcheva, and
  Ruszczynski}]{shapiro2021lectures}
Shapiro, A., Dentcheva, D., Ruszczynski, A., 2021. Lectures on stochastic
  programming: modeling and theory. SIAM.

\bibitem[{Shehadeh(2020)}]{DMFRS}
Shehadeh, K.~S., 2020. Distributionally robust optimization approaches for a
  stochastic mobile facility routing and scheduling problem. arXiv preprint
  arXiv:2009.10894.

\bibitem[{Shehadeh and Sanci(2021)}]{ShehadehSanci}
Shehadeh, K.~S., Sanci, E., 2021. Distributionally robust facility location
  with bimodal random demand. Computer and Operations Research.

\bibitem[{Shehadeh and Tucker(2020)}]{shehadeh2020distributionallyTucker}
Shehadeh, K.~S., Tucker, E.~L., 2020. A distributionally robust optimization
  approach for location and inventory prepositioning of disaster relief
  supplies. arXiv preprint arXiv:2012.05387.

\bibitem[{Shin and Lee(2018)}]{shin2018improving}
Shin, K., Lee, T., 2018. Improving the measurement of the korean emergency
  medical system's spatial accessibility. Applied geography 100, 30--38.

\bibitem[{Shishebori and Babadi(2015)}]{shishebori2015robust}
Shishebori, D., Babadi, A.~Y., 2015. Robust and reliable medical services
  network design under uncertain environment and system disruptions.
  Transportation Research Part E: Logistics and Transportation Review 77,
  268--288.

\bibitem[{Silva and Serra(2008)}]{silva2008locating}
Silva, F., Serra, D., 2008. Locating emergency services with different
  priorities: the priority queuing covering location problem. Journal of the
  Operational Research Society 59~(9), 1229--1238.

\bibitem[{Smith et~al.(2013)Smith, Harper, and Potts}]{smith2013bicriteria}
Smith, H.~K., Harper, P.~R., Potts, C.~N., 2013. Bicriteria efficiency/equity
  hierarchical location models for public service application. Journal of the
  Operational Research Society 64~(4), 500--512.

\bibitem[{Smith and Winkler(2006)}]{smith2006optimizer}
Smith, J.~E., Winkler, R.~L., 2006. The optimizer’s curse: Skepticism and
  postdecision surprise in decision analysis. Management Science 52~(3),
  311--322.

\bibitem[{Snyder(2006)}]{snyder2006facility}
Snyder, L.~V., 2006. Facility location under uncertainty: a review. IIE
  transactions 38~(7), 547--564.

\bibitem[{Sommers(2015)}]{sommers2015health}
Sommers, B.~D., 2015. Health care reform's unfinished work—remaining barriers
  to coverage and access. New England Journal of Medicine.

\bibitem[{Song et~al.(2013)Song, Hill, Bennet, Vavasis, and
  Oriol}]{song2013mobile}
Song, Z., Hill, C., Bennet, J., Vavasis, A., Oriol, N.~E., 2013. Mobile clinic
  in massachusetts associated with cost savings from lowering blood pressure
  and emergency department use. Health affairs 32~(1), 36--44.

\bibitem[{Sorensen and Church(2010)}]{sorensen2010integrating}
Sorensen, P., Church, R., 2010. Integrating expected coverage and local
  reliability for emergency medical services location problems. Socio-Economic
  Planning Sciences 44~(1), 8--18.

\bibitem[{Vidyarthi and Kuzgunkaya(2015)}]{vidyarthi2015impact}
Vidyarthi, N., Kuzgunkaya, O., 2015. The impact of directed choice on the
  design of preventive healthcare facility network under congestion. Health
  care management science 18~(4), 459--474.

\bibitem[{Vilkkumaa and Liesi{\"o}(2021)}]{vilkkumaa2021causes}
Vilkkumaa, E., Liesi{\"o}, J., 2021. What causes post-decision disappointment?
  estimating the contributions of systematic and selection biases. European
  Journal of Operational Research.

\bibitem[{Wang(2012)}]{wang2012measurement}
Wang, F., 2012. Measurement, optimization, and impact of health care
  accessibility: a methodological review. Annals of the Association of American
  Geographers 102~(5), 1104--1112.

\bibitem[{Wang et~al.(2020)Wang, Chen, and Liu}]{wang2020distributionally}
Wang, S., Chen, Z., Liu, T., 2020. Distributionally robust hub location.
  Transportation Science 54~(5), 1189--1210.

\bibitem[{Wang et~al.(2021)Wang, Yang, Yang, and Gao}]{wang2021two}
Wang, W., Yang, K., Yang, L., Gao, Z., 2021. Two-stage distributionally robust
  programming based on worst-case mean-cvar criterion and application to
  disaster relief management. Transportation Research Part E: Logistics and
  Transportation Review 149, 102332.

\bibitem[{Wu et~al.(2015)Wu, Du, and Xu}]{wu2015approximation}
Wu, C., Du, D., Xu, D., 2015. An approximation algorithm for the two-stage
  distributionally robust facility location problem. In: Advances in Global
  Optimization. Springer, pp. 99--107.

\bibitem[{Yin et~al.(2018)Yin, He, Liu, Chen, and Gao}]{yin2018inequality}
Yin, C., He, Q., Liu, Y., Chen, W., Gao, Y., 2018. Inequality of public health
  and its role in spatial accessibility to medical facilities in china. Applied
  Geography 92, 50--62.

\bibitem[{Yoon et~al.(2021)Yoon, Albert, and White}]{yoon2021stochastic}
Yoon, S., Albert, L.~A., White, V.~M., 2021. A stochastic programming approach
  for locating and dispatching two types of ambulances. Transportation Science
  55~(2), 275--296.

\bibitem[{Zahiri et~al.(2014{\natexlab{a}})Zahiri, Tavakkoli-Moghaddam,
  Mohammadi, and Jula}]{zahiri2014multi}
Zahiri, B., Tavakkoli-Moghaddam, R., Mohammadi, M., Jula, P.,
  2014{\natexlab{a}}. Multi-objective design of an organ transplant network
  under uncertainty. Transportation Research Part E: Logistics and
  Transportation Review 72, 101--124.

\bibitem[{Zahiri et~al.(2014{\natexlab{b}})Zahiri, Tavakkoli-Moghaddam, and
  Pishvaee}]{zahiri2014robust}
Zahiri, B., Tavakkoli-Moghaddam, R., Pishvaee, M.~S., 2014{\natexlab{b}}. A
  robust possibilistic programming approach to multi-period
  location--allocation of organ transplant centers under uncertainty. Computers
  \& Industrial Engineering 74, 139--148.

\bibitem[{Zhang et~al.(2010)Zhang, Berman, Marcotte, and
  Verter}]{zhang2010bilevel}
Zhang, Y., Berman, O., Marcotte, P., Verter, V., 2010. A bilevel model for
  preventive healthcare facility network design with congestion. IIE
  Transactions 42~(12), 865--880.

\bibitem[{Zhang et~al.(2009)Zhang, Berman, and Verter}]{zhang2009incorporating}
Zhang, Y., Berman, O., Verter, V., 2009. Incorporating congestion in preventive
  healthcare facility network design. European Journal of Operational Research
  198~(3), 922--935.

\bibitem[{Zhang and Li(2015)}]{zhang2015novel}
Zhang, Z.-H., Li, K., 2015. A novel probabilistic formulation for locating and
  sizing emergency medical service stations. Annals of Operations Research
  229~(1), 813--835.

\end{thebibliography}
\end{document}